\documentclass[12pt, draftclsnofoot, onecolumn]{IEEEtran}
\makeatletter
\def\ps@headings{%
\def\@oddhead{\mbox{}\scriptsize\rightmark \hfil \thepage}%
\def\@evenhead{\scriptsize\thepage \hfil \leftmark\mbox{}}%
\def\@oddfoot{}%
\def\@evenfoot{}}
\makeatother
\usepackage{xcolor}
\usepackage{graphicx, subfigure, epsfig}
\usepackage{url}
\usepackage[ruled,vlined]{algorithm2e}
\usepackage{algpseudocode}

\usepackage{tikz}
\usepackage{listings}
\usepackage{amssymb,amsfonts}
\usepackage[cmex10]{amsmath}
\usepackage{color}
    \definecolor{lightgray}{rgb}{0.95, 0.95, 0.95}
    \definecolor{darkgray}{rgb}{0.4, 0.4, 0.4}
    \definecolor{purple}{rgb}{0.65, 0.12, 0.82}
    \definecolor{ocherCode}{rgb}{1, 0.5, 0} 
    \definecolor{blueCode}{rgb}{0, 0, 0.93} 
    \definecolor{greenCode}{rgb}{0, 0.6, 0} 
\usepackage{upquote}
\usepackage{algpseudocode}
\usepackage[square, comma, sort&compress, numbers]{natbib} 
\usepackage{diagbox}
\usepackage{multirow}
\usepackage{booktabs}
\usepackage{tabu}

\newcommand {\beq} {\begin{equation}}
\newcommand {\eeq} {\end{equation}}

\newtheorem{theorem}{\bf Theorem}
\newtheorem{definition}{\bf Definition}
\newcommand{\RNum}[1]{\uppercase\expandafter{\romannumeral #1\relax}}
\newcommand{\tabincell}[2]{\begin{tabular}{@{}#1@{}}#2\end{tabular}}

\renewcommand\thesection{\arabic{section}}
\renewcommand\thetable{\arabic{table}}
\newcommand\copyrighttext{
	\footnotesize \textcopyright This work has been submitted to the IEEE for possible publication. Copyright may be transferred without notice, after which this version may no longer be accessible.}
\newcommand\copyrightnotice{
	\begin{tikzpicture}[remember picture,overlay]
	\node[anchor=south,yshift=10pt] at (current page.south) {\fbox{\parbox{\dimexpr\textwidth-\fboxsep-\fboxrule\relax}{\copyrighttext}}};
	\end{tikzpicture}
}

\begin{document}

\title{IoTAthena: Unveiling IoT Device Activities
from Network Traffic}
\author{\IEEEauthorblockN{
Yinxin~Wan,~\IEEEmembership{Student Member,~IEEE,}
Kuai~Xu,~\IEEEmembership{Senior Member,~IEEE,}
Feng~Wang,~\IEEEmembership{Member,~IEEE,}
Guoliang~Xue,~\IEEEmembership{Fellow,~IEEE}
\thanks{}
\thanks{Wan, Xu, Wang, and Xue are all affiliated with Arizona State University. Emails: \{ywan28; kuai.xu; fwang25; xue\}@asu.edu.
This research was supported in part by NSF grants 1816995, 1717197, and 1704092.
The information reported here does not reflect the position or the policy of the funding agency.
}
}
}
\maketitle

\copyrightnotice
\begin{abstract}
The recent spate of cyber attacks towards Internet of Things (IoT) devices in smart homes calls for effective techniques to understand, characterize, and unveil IoT device activities.
In this paper, we present a new system, named IoTAthena, to unveil IoT device activities from raw network traffic consisting of timestamped IP packets.
IoTAthena characterizes each IoT device activity using an activity signature consisting of an ordered sequence of IP packets with inter-packet time intervals.
IoTAthena has two novel polynomial time algorithms, {\tt sigMatch} and {\tt actExtract}.
For any given signature, {\tt sigMatch} can capture all matches of the signature in the raw network traffic.
Using {\tt sigMatch} as a subfunction, {\tt actExtract} can accurately unveil the sequence of various IoT device activities from the raw network traffic.
Using the network traffic of heterogeneous IoT devices collected at the router of a real-world smart home testbed and a public IoT dataset, we demonstrate that IoTAthena is able to characterize and generate activity signatures of IoT device activities and accurately unveil the sequence of IoT device activities from raw network traffic.

\end{abstract}

\section{Introduction}
\label{sec:01}
\noindent
In today's smart homes, various IoT devices can connect to the Internet via home routers with wired cable connections or wireless communications such as WiFi, Bluetooth, and ZigBee.
The proliferating IoT devices in smart homes bring many innovative applications and services such as improved home automation and safety, efficient energy, and connected healthcare.
However, the recent spate of cyber attacks and threats~\cite{Antonakakis:Security17, Kambourakis:MILCOM17,Herwig:NDSS19,Ronen:SP17,Ronen:EuroSP16,Morgner:WiSec17,Sivaraman:WiSec16,Ryan:Blackhat13,Zillner:Blackhat15,Demetriou:WiSec2017} towards a wide range of IoT devices with flawed system designs and weak security management calls for effective techniques to understand, characterize, and unveil the detailed activities of heterogeneous IoT devices, e.g., {\it when} and {\it how} a smart lock is opened.

In the research literature, there have been some prior studies on characterizing behavioral patterns of IoT devices and identifying IoT device types and activities using supervised machine learning models or the simple request/reply pattern matching~\cite{Miettinen:ICDCS17,Bezawada:ASHES18,Sivanathan:TMC18,Xu:IWQoS19,Wan:TNSE20,Wan:INFOCOM20,OConnor:WiSec2019,Acar:WiSec2020,Trimananda:NDSS20,Zhang:CCS18,Jia:InfoCom2018}.
However, little effort has been devoted to the understanding and characterization of full and detailed signatures of IoT device activities which shed light on when and how IoT devices communicate with cloud servers and smartphones for carrying out what device activities.

This paper presents IoTAthena, a system to generate detailed signatures of IoT device activities from IoT network traffic and to unveil IoT device activities via time-sensitive subsequence matching.
IoTAthena first collects the background traffic of IoT devices and the normal network traffic triggered by IoT device activities with packet capturing tools
on programmable home routers.
It then characterizes each IoT device activity using an activity signature consisting of an ordered sequence of IP packets with inter-packet time intervals\footnote{An inter-packet time interval is calculated from the timestamp difference of two consecutive IP packets in the packet sequence.}.

Our IoT device activity signature generation is inspired by PingPong~\cite{Trimananda:NDSS20}, which generates packet-level signatures of IoT device activities in the form of abstracted packet-pairs with ping/pong, i.e., request/reply, patterns.
It has been demonstrated~\cite{Trimananda:NDSS20} that many IoT device activities can be efficiently captured with the use of ping/pong like signatures.
However, such short signatures have their limitations.
For example, our experiments of running the PingPong open source package were unable to generate signatures of the WiFi and Bluetooth locking or unlocking activities of August Lock, as well as the autolocking activity.
In contrast, IoTAthena's detailed activity signature carries crucial information for characterizing more IoT device activities than~\cite{Trimananda:NDSS20} and differentiating IoT device activities with overlapping packet pairs or packet sequences.

To unveil IoT device activities from network traffic logs, IoTAthena relies on two novel algorithms, {\tt sigMatch} and {\tt actExtract}.
The {\tt sigMatch} algorithm can effectively capture all matches of a given activity signature from the network traffic log in polynomial time.
Using {\tt sigMatch} as a subfunction, the {\tt actExtract} algorithm can accurately unveil the sequence of IoT device activities from raw network traffic logs, also in polynomial time.
Our experimental evaluations of IoTAthena were based on $16$ IoT devices in a real-world smart home environment, and a public IoT dataset of $25$ IoT devices~\cite{Ren:IMC19}. Our experimental results showed that IoTAthena can effectively generate the detailed signatures of IoT device activities and accurately unveil future activities of these IoT devices.

The main contributions of this paper are the following:
\begin{itemize}
\item
We develop a systematic approach to programmatically generate detailed signatures of IoT device activities consisting of an ordered sequence of IP packets with inter-packet time intervals.

\item
We design two novel polynomial time algorithms, {\tt sigMatch} and {\tt actExtract}, for capturing all matches of a given IoT device activity signature and unveiling activity sequences of all IoT devices from the network traffic logs.

\item
We conduct extensive experiments using a smart home testbed and a public IoT dataset~\cite{Ren:IMC19} to demonstrate that IoTAthena can accurately capture the activities of a wide range of heterogeneous IoT devices.
\end{itemize}

The remainder of this paper is organized as follows.
We first discuss the related work in Section~\ref{sec:related}.
In Section~\ref{sec:system}, we present an overview of the IoTAthena system.
In Section~\ref{sec:traffic}, we describe how IoTAthena collects IoT network traffic, and analyzes and characterizes the background traffic of IoT devices.
In Section~\ref{sec:sig}, we formally define the IoT device activity signatures, and describe the process of generating the signature for each device activity.
In Section~\ref{sec:alg}, we present the {\tt sigMatch} and {\tt actExtract} algorithms with theoretical analysis.
In Section~\ref{sec:eval}, we present our experimental evaluation results.
Section~\ref{sec:conc} concludes the paper and outlines our future work.

\section{Related Work}
\label{sec:related}
\noindent
The recent growth and deployment of IoT devices in smart homes have attracted the networking research community to study the traffic characterization and behavioral fingerprinting of IoT devices, and explore network traffic to discover IoT devices' types and activities.
Most of the existing studies~\cite{Miettinen:ICDCS17,Xu:IWQoS19,Wan:TNSE20,Ren:IMC19,Gao:DSN,Siby:IoTPTS17,Gu:MASS18,Jafari:MILCOM18,Ma:INFOCOM20,Gu:INFOCOM20,Huang:TWC20,Zhang:TWC20,Zou:TWC17} in IoT traffic characterization and fingerprinting are interested in a wide range of traffic features from TCP/IP protocols as well as from IoT wireless communication channels.
For example,~\cite{Huang:TWC20} utilizes the wireless radio propagation patterns of IoT devices for secure authentication.
\cite{Zou:TWC17,Zhang:TWC20} explore the captured WiFi signals in home network for applications of localization and positioning. 
\cite{Xu:IWQoS19,Wan:TNSE20}, on the other hand, examine the home network traffic at flow level to model and profile IoT devices.
These prior research provide critical insights for understanding traffic patterns of heterogeneous IoT devices and identifying IoT device models or types for IoT device discovery and management, IoT application performance monitoring, and vulnerability and security analysis.

In light of the recent IoT Botnets exploiting and control thousands of vulnerable IoT devices~\cite{Antonakakis:Security17,Alrawi:SP19,Kumar:Security19,Fernandes:SP16,Fernandes:SP17,Mosenia:TETC16,Morgner:SP20,Yu:Security20,Herwig:NDSS19}, some research efforts have proposed innovative methods of classifying IoT devices based on machine learning, statistical inference, or passive traffic measurement~\cite{Miettinen:ICDCS17,Bezawada:ASHES18,Sivanathan:TMC18}.
For example, the IoTSentinel system~\cite{Miettinen:ICDCS17} first extracts $23$ traffic features of IoT network traffic, and subsequently builds Random Forest classifiers to identify IoT device types.
Similarly, IoTSense~\cite{Bezawada:ASHES18} fingerprints the behaviors of IoT device types with feature vectors from packet headers and payload, and builds several machine learning classifiers for effectively detecting IoT device types based on the trained behavioral fingerprinting. The research in~\cite{Sivanathan:TMC18} first monitors a smart IoT environment with various IoT devices for six months for extensive IoT network traffic analysis, and then builds a machine learning framework for classifying IoT device types.

As homeowners continue to deploy smart home IoT devices such as smart locks and security cameras for mission-critical applications, accurately identifying IoT device activities via supervised machine learning models~\cite{OConnor:WiSec2019,Acar:WiSec2020} and deterministic inference~\cite{Trimananda:NDSS20,Zhang:CCS18} becomes an urgent research problem.
For example, HomeSnitch~\cite{OConnor:WiSec2019} constructs bidirectional application data unit exchanges for representing IoT application behaviors and applies supervised machine learning classifiers to classify IoT application behaviors and identifying unknown behaviors.
Similarly, Peek-a-Boo~\cite{Acar:WiSec2020} demonstrates the feasibility of identifying the types, states, and IoT devices' activities via machine learning techniques from an attacker's perspective.
The closest work to ours is PingPong~\cite{Trimananda:NDSS20}, which explores the sequential and directional ``ping/pong'' behavioral patterns between cloud servers and IoT devices or between cloud servers and smartphones.
The experiments in~\cite{Trimananda:NDSS20} have shown that the simple ping/pong packet-pairs with payload size and traffic directions can effectively detect many IoT devices' activities.
HoMonit~\cite{Zhang:CCS18}, another work of deterministically detecting IoT device activity, monitors encrypted wireless traffic of some home apps and infers smart app activities based on the deterministic finite automaton (DFA) model of smart app behavior and wireless side-channel analysis.
IoTGaze~\cite{Gu:INFOCOM20} also builds up a system to identify IoT device activities using the sniffed wireless traffic.

Inspired and motivated by these studies on identifying IoT device types and/or activities, our proposed IoTAthena system is focused on understanding traffic signatures of IoT device activities and accurately extracting device activities from IoT network traffic.
The insights from the unveiled IoT device activities have a broad range of applications such as anomaly detection, e.g., an unauthorized user is watching the video stream of the surveillance camera, IoT device malfunction detection, e.g., a smart plug shows two consecutive {\it on} activities, and smart home safety, e.g., the smart lock was unlocked remotely by an unauthorized user.

Note that our work is significantly different from~\cite{Miettinen:ICDCS17,Bezawada:ASHES18,Sivanathan:TMC18} in the way that our objective is generating signatures for concrete IoT device activities such as {\it on} or {\it off} activities of a smart plug and unveiling these activities from network traffic, instead of identifying IoT device models or types.
Different from machine learning based solutions~\cite{OConnor:WiSec2019,Acar:WiSec2020}, IoTAthena adopts a {\it white-box} approach to programmatically generate activity signatures of IoT device activities consisting of ordered sequences of IP data packets with relative timestamps.
IoTAthena's signature generation module is inspired by PingPong~\cite{Trimananda:NDSS20}, but it generates a {\it full} signature for each IoT device activity and introduces a novel time-sensitive subsequence matching approach for unveiling IoT device activities from new IoT network traffic logs.

\section{IoTAthena System Overview}
\label{sec:system}
\noindent
Developing effective techniques to understand and report IoT device activities, {\em e.g., the smart lock of the home's main entrance is unlocked remotely with a smartphone app}, is crucial for ensuring the physical and property safety of these devices' homeowners.
Our real-world experiments with August Lock and other IoT devices demonstrated the feasibility of developing an automated system to learn and generate signatures of IoT device activities and use them for unveiling IoT device activities from network traffic logs.
Such a system is urgently needed for understanding what is happening to IoT devices in millions of smart homes and for detecting suspicious and unauthorized behaviors towards critical home devices.

\begin{figure*}[htbp]
\centering
\includegraphics[width=0.95\textwidth]{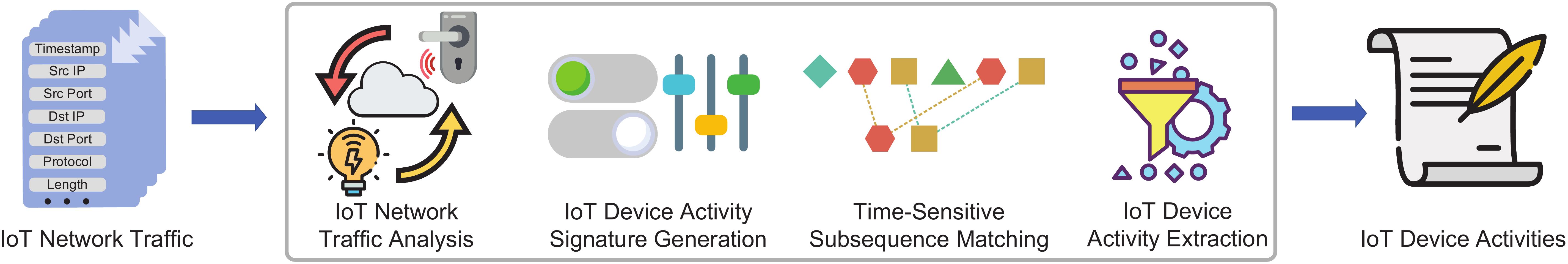}
\caption{Overall architecture of the IoTAthena system for
unveiling IoT device activities from IoT network traffic.}
\label{fig:sys}
\end{figure*}

In this paper, we propose a new system, named IoTAthena, to automatically and accurately unveil IoT device activities from smart home network traffic logs.
Fig.~\ref{fig:sys} illustrates the overall architecture of IoTAthena, which includes four key system modules:
i) IoT network traffic analysis,
ii) IoT device activity signature generation,
iii) time-sensitive subsequence matching, and
iv) IoT device activity extraction.

The {\it IoT network traffic analysis} module takes IoT network traffic during the intentionally ``silent'' period, and characterizes background network traffic for each IoT device.
The {\it IoT device activity signature generation} module collects the corresponding network traffic of each IoT device activity by intentionally triggering the activity and collecting the traffic.
The collected IoT network traffic along with the labelled activity logs serve as the ground truth for generating the signature of each IoT device activity consisting of an ordered sequence of IP packets with inter-packet time intervals.
The {\it time-sensitive subsequence matching} module relies on the {\tt sigMatch} algorithm to capture all matches of each IoT device activity signature in the network traffic log, while the {\it IoT device activity extraction} module relies on {\tt actExtract} to unveil the sequence of IoT device activities from the network traffic log.

In summary, IoTAthena adopts a white-box approach to first generate signatures of IoT device activities consisting of ordered sequences of IP packets with inter-packet time interval information. 
Subsequently, IoTAthena applies efficient matching algorithms for deterministically unveiling the sequence of IoT device activities from the network traffic log, unlike black-box machine learning classification models~\cite{Miettinen:ICDCS17,Bezawada:ASHES18,Sivanathan:TMC18,Acar:WiSec2020}.

\section{Network Traffic Collection and Analysis}
\label{sec:traffic}
\noindent
Network traffic of IoT devices embeds rich information on device types and their behavioral patterns~\cite{Ren:IMC19,Sivanathan:TMC18}.
In this section, we describe how to collect and analyze IoT network traffic in order to characterize and generate signatures of IoT device activities.

\subsection{IoT Network Traffic Collection}
\label{sec:traffic-A}
\begin{figure}[htbp]
\centering
\includegraphics[width=3.5in]{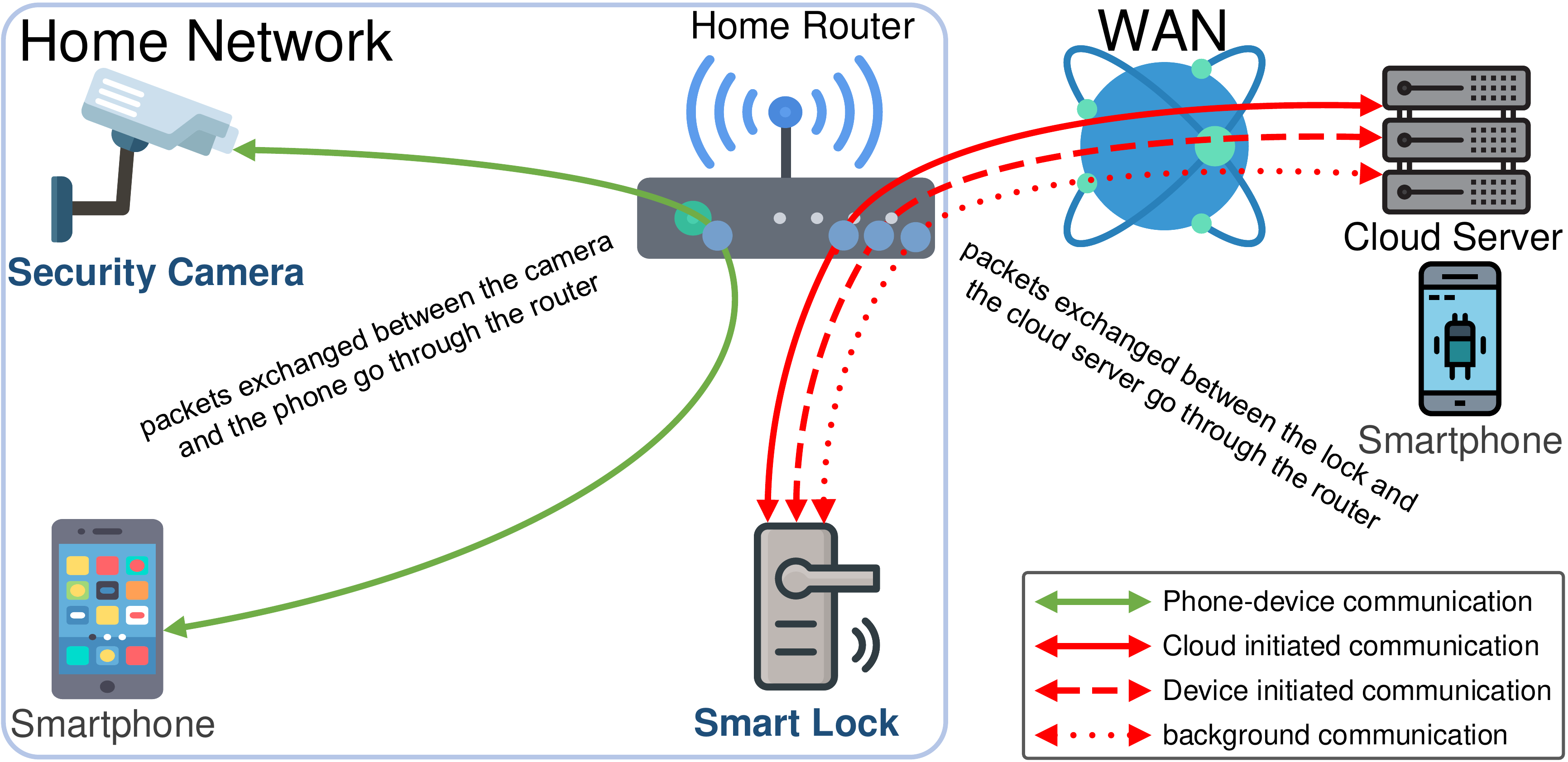}
\caption{Illustration of data flows initiated from or destined to
IoT devices, using smart lock and security camera as examples.}
\vspace{-0.1in}
\label{fig:control}
\end{figure}

\noindent
Fig.~\ref{fig:control} illustrates the data flows initiated from an IoT device or destined to an IoT device in a smart home environment.
For clarity, we use two IoT devices as examples: a smart lock and a security camera.
A user usually interacts with an IoT device using the device's companion app on the smartphone in the home or outside the home, e.g., in the office or on the road.
The app first communicates with the cloud server which in turn generates traffic between the cloud server and the device, as illustrated by the solid red line between the smart lock and the cloud server.
Sometimes, the smartphone directly communicates with the device without involving the cloud server, such as streaming request on the security camera, illustrated by the solid green line between the smart phone and the security camera.

The user can also manually operate the device in the traditional way, such as locking the smart lock manually.
This action causes the device to update its status to the cloud server immediately following the action.
In addition, the user can communicate directly with the device locally through a non-WiFi communication channel, such as Bluetooth or ultrawideband (UWB) when the user is in the vicinity of the device.
This action also causes device initiated status updates.
Furthermore, automatic device operations such as the smart lock's autolocking function also introduce status update traffic.
These types of device initiated communications are illustrated by the dashed red line.
There also exists traffic introduced by device background operations such as device firmware update checks.
We use the dotted red line between the smart lock and the cloud server to illustrate these data flows.

IoTAthena collects the network traffic at the programmable home router, which enables the capture of incoming and outgoing packets of all above mentioned device-related operations. 
The smart home router is a desirable centralized location for data collection, considering its switching and routing function, sufficient computational and processing capacities, and the design transparency to IoT devices and apps.
In our experiments, we used Linksys WRT1900AC WiFi home routers which run the open-source Linux-based OpenWrt operating system for network traffic collection. The data processing and analysis were performed offline in this study. 
One of our future work is designing and implementing a prototype system on commodity home routers to evaluate the real-time feasibility of IoTAthena for unveiling IoT device activities on the fly. 

\subsection{IoT Network Traffic Analysis}
\label{sec:backgroundnoise}
\noindent
The network traffic collected at the home router can be classified into two parts: the first part consists of traffic between the IoT devices and the cloud servers, while the second part consists of internal LAN traffic such as address resolution protocol (ARP) requests and simple service discovery protocol (SSDP) broadcast packets.
In order to separate the logs of an individual IoT device from the mixed home network traffic, IoTAthena first identifies each device's unique IP address via the mapping of its media access control (MAC) address and host name in the dynamic host configuration protocol (DHCP) packets.
It subsequently uses the device IP address as the unique {\it cluster} key to separate IoT network traffic into individual traffic clusters to simplify further analysis.

We carefully studied the network traffic within each individual traffic cluster of an IoT device.
We can clearly observe the network traffic one would anticipate for normal IoT device activities, e.g., users issuing locking or unlocking commands for August Lock via the smartphone app.
Surprisingly, we also discovered a significant amount of network traffic when there is no human-triggered or environment-triggered activity.
We use the term {\em background traffic} to denote such network traffic, i.e., network traffic not triggered by human or environment.
In order to gain a thorough understanding of IoT background traffic, we left the devices in our controlled smart home environment without any human interactions for one week and consider the network traffic cluster of each IoT device during this ``silent'' period as background traffic.
By separating IP data packets based on the destination (and source) ports of the outgoing (and incoming) traffic, we observed that these IoT devices typically exchange messages with the remote cloud servers on the well-known application ports such as 22/TCP (SSH), 53/UDP (DNS), 80/TCP (HTTP), 123/UDP (NTP), 5353/UDP (mDNS). 
This observation leads us to classify IoT background traffic into three categories: management and service, signal and update, and random noise.

The management and service traffic is mainly used to manage and maintain the devices, e.g., periodical time synchronizations with NTP servers.
The signal and update traffic corresponds to keep-alive signals and regular firewall update checks between IoT devices and cloud servers.
The random noise traffic is mostly generated by other IoT or non-IoT devices in the local home network for a variety of reasons, e.g., ARP requests, SSDP broadcasts, and multicast DNS (mDNS) traffic from Apple Bonjour protocol for automatic device and service discovery.

The background traffic analysis not only removes unnecessary noise for characterizing and generating signatures of IoT device activities, but also sheds light on the potential vulnerabilities of the protocol stacks of mission-critical IoT devices in millions of smart homes.
For example, our analysis discovered the usage of non-encrypted and insecure Telnet and HTTP sessions between some camera devices and cloud servers, for logins and firmware update checks.
Discovering and mitigating security weaknesses of IoT devices is beyond the scope of this paper.

\section{IoT Device Activity Signatures}
\label{sec:sig}
\begin{figure*}[ht]
\centering
\subfigure[August Lock {\it WiFi (un)locking}.]{
    \includegraphics[width=3.0in]{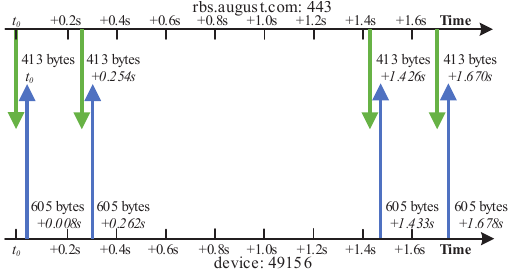}
}
\hspace{0.05in}
\subfigure[August Lock {\it Bluetooth (un)locking}.]{
    \includegraphics[width=3.0in]{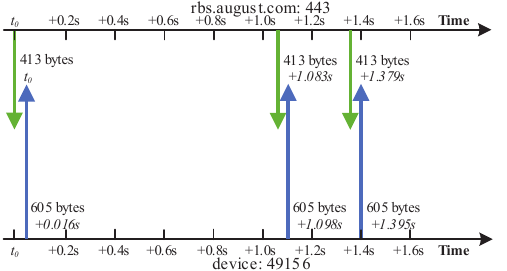}
}
\vspace{-0.10in}
\caption{Packet sequences of August Lock's activities:
the pattern in (b)
{\em seems like}
a subsequence of the pattern in (a).}
\label{fig:locksig}
\vspace{-0.15in}
\end{figure*}

\noindent
Consistent with the findings of PingPong~\cite{Trimananda:NDSS20}, we observed repetitive network packet sequences that correspond to repeated device activities in the network traffic collected at the router of the smart home network.
We also observed certain August Lock activities resulting packet sequences that are challenging for PingPong to recognize.
Fig.~\ref{fig:locksig} illustrates such an example.
The Bluetooth (un)locking\footnote{The locking activity and the unlocking activity exhibit the same packet sequences and inter-packet time intervals because of the simple lock/unlock state transitions.
The encrypted application data prevents us for further differentiating these two activities with network traffic only.
We generate a unique signature for each {\em indistinguishable activity group}.
For example, we use {\tt (un)locking} for short to denote either the {\tt locking} activity or the {\tt unlocking} activity.
Similarly, we use {\tt on or off} to denote either the {\tt on} activity or the {\tt off} activity.} activity's packet sequence ($3$ pairs as illustrated in Fig.~\ref{fig:locksig}(b)) is a subset of the WiFi (un)locking activity's packet sequence ($4$ pairs as illustrated in Fig.~\ref{fig:locksig}(a)).
The clustering of re-occurring packet pairs approach in PingPong cannot distinguish WiFi (un)locking from Bluetooth (un)locking.
In fact, it is difficult to distinguish these two activities in the network traffic solely based on request/reply patterns, which leads us to consider more information (the full detailed packet sequence) and inter-packet time intervals to characterize IoT device activities. 
The time intervals between consecutive packets provide critical information to effectively and accurately differentiate IoT device activities such as those in Fig.~\ref{fig:locksig} that share overlapping packet sequences and happen very closely in time.

\subsection{Inter-Packet Time Interval Measurement}
\label{sec:interval}
\noindent
Because IoTAthena collects network traffic at the home router, the inter-packet time interval is essentially the round-trip time (RTT) between the home router and IoT devices in the smart home plus the processing time at the device (LAN).
The time interval could also be the RTT between the home router and cloud servers across the Internet plus the processing time at the cloud server (WAN).

\begin{figure}[htbp]
\centering
\includegraphics[width=6.4in]{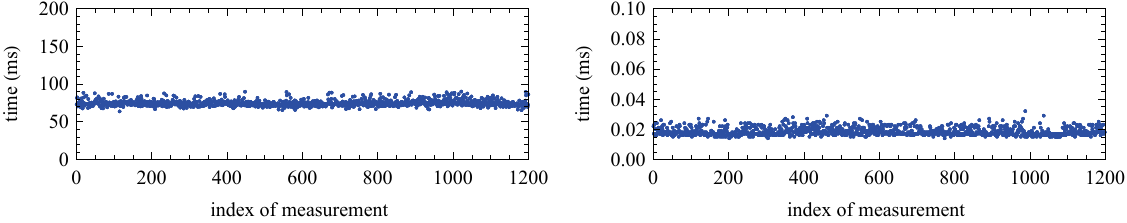}
\vspace{-0.10in}
\caption{Inter-packet time intervals:
large values in WAN (left plot) vs small values in LAN (right plot).}
\label{fig:interpkttime}
\vspace{-0.10in}
\end{figure}

Fig.~\ref{fig:interpkttime} illustrates the time interval between the first and second packets (left plot), and the time interval between the second and third packets (right plot), of $1,200$ repeated {\it on} activities of TP-Link Plug over a $24$-hour span.
We observe that {\em the inter-packet time intervals exhibit stable and consistent patterns, with small variances}.
However, {\em the time interval between one pair of consecutive packets may significantly differ from that between another pair of consecutive packets}.
Specifically, the interval between the first and second packets has a mean ($\mu$) of $76.73ms$ and a standard deviation ($\sigma$) of $0.003995ms$, while the interval between the second and third packets has a mean ($\mu$) of $0.02ms$ and a standard deviation ($\sigma$) of $0.000003ms$ for TP-Link Plug's {\it on} activity.
The unstable wireless channel between the IoT devices and the home router could result in packet loss and retransmission which contribute to the fluctuation in the LAN inter-packet time intervals. 
The uncertain number of retransmissions in the MAC layer affects the inter-packet time intervals in our collected traffic log. 
However, compared with the short wireless transmission delay, the local processing time at the IoT device still dominates the LAN inter-packet time intervals, as we observe from the right plot in Fig.~\ref{fig:interpkttime}.
These key observations inspire us to include inter-packet time intervals as an important component in characterizing the signatures of IoT device activities.

\subsection{IoT Device Activity Signature Definition}
\label{sec:signaturedef}
\noindent
A network data packet $p$ collected at the home router is an $8$-tuple, where the first through eighth fields are {\em timestamp}, {\em IoT device internal IP address}, {\em canonical remote cloud server name}, {\em remote application port}, {\em protocol}, {\em traffic direction}, {\em packet length}, and {\em application-layer data}, respectively.
It is important to note that each TCP/IP data packet carries a variety of traffic features including those in the 8-tuple. 
However, this study only selects the features that provide additional information on identifying and differentiating IoT device activities, while skipping the features, e.g., Time to Live (TTL), sequence and acknowledgement numbers with redundant or little contributions towards device activity identification.

%
We use $p.t$ to denote the timestamp of packet $p$,
and use $\widehat{p}$ to denote the $7$-tuple obtained by deleting the
first field (timestamp) in $p$.
We call $\widehat{p}$ the {\em base packet} of packet $p$.

\textcolor{black}{
\begin{definition}
\label{def:sig}
The signature of an IoT device activity is given by
an ordered sequence of $n$ base packets
$(\widehat{q_1}, \widehat{q_2}, \ldots, \widehat{q_n})$,
together with an ordered sequence of $n-1$ inter-packet time intervals
$(\tau_1, \tau_2, \ldots, \tau_{n-1})$,
where $\tau_j > 0$ is the time interval between the $j$th packet and the
$j+1$th packet, $j=1, 2, \ldots, n-1$.
The number of base packets, $n$, in each device activity signature
is determined by the observed TCP/IP data packets triggered by the activity
minus the protocol-specific packets, e.g., TCP three-way handshake,
and the regular heart-beat signals between the device and the remote cloud server.
\hfill$\Box$
\end{definition}
}

Instead of using $(\widehat{q_1}, \widehat{q_2}, \ldots, \widehat{q_n})$ and $(\tau_1, \tau_2, \ldots, \tau_{n-1})$ to represent a signature, we can equivalently represent the same signature using a sequence of $n$ packets $(\rho_1, \rho_2, \ldots, \rho_n)$, where $\widehat{\rho_j} = \widehat{q_j}$ for $j=1, 2, \ldots, n$ and $\rho_{j+1}.t - \rho_{j}.t = \tau_j$ for $j=1, 2, \ldots, n-1$.
In this representation, the time interval between the $j$th packet and the $j+1$th packet can be uniquely computed by $\tau_j = \rho_{j+1}.t - \rho_{j}.t$.

The signature of an IoT device activity is a constant, as defined in {\bf Definition}~\ref{def:sig}.
The above {\em alternative} representation of the signature, however, does not look like a constant {\em in format}.
For example, for any given real number $c$, $(p_1, p_2, \ldots, p_n)$ and $(\rho_1, \rho_2, \ldots, \rho_n)$ denote exactly the same signature, provided that $\widehat{p_j} = \widehat{\rho_j}$, $p_{j}.t = \rho_{j}.t + c$, for $j=1, 2, \ldots, n$.
Since $(p_1, p_2, \ldots, p_n)$ and $(\rho_1, \rho_2, \ldots, \rho_n)$ define exactly the same sequence of $n$ base packets $(\widehat{p_1}, \widehat{p_2}, \ldots, \widehat{p_n})$ $=$ $(\widehat{\rho_1}, \widehat{\rho_2}, \ldots, \widehat{\rho_n})$ and exactly the same sequence of $n-1$ inter-packet time intervals $(p_2.t - p_1.t, p_3.t - p_2.t, \ldots, p_n.t - p_{n-1}.t)$ $=$ $(\rho_2.t - \rho_1.t, \rho_3.t - \rho_2.t, \ldots, \rho_n.t - \rho_{n-1}.t)$, {\em we can use this alternative representation without losing any accuracy.}

Given the above discussions, we will denote a signature of an IoT device activity using an ordered sequence of packets $(q_1, q_2, \ldots, q_n)$, where the timestamp fields are only used to compute the inter-packet time intervals $\tau_j = q_{j+1}.t - q_{j}.t$.
For this reason, we also call the timestamp fields in a signature {\em relative timestamps}.
We set $q_1.t$ to $0$ for simplicity.

\subsection{Automated Signature Generation}
\label{sec:sig-gen}
\noindent
Towards automatically generating activity signatures of IoT device activities, we first follow the same practice as~\cite{Zhu:TNSE20} to compile a complete list of any given IoT device's activities from the AndroidManifest.xml file of the device's companion app.
We then write scripts using command-line tool and scripting feature in Android Debug Bridge (ADB) to automate the user interactions with IoT devices such as turning on/off Philips Hue and (un)locking of August Lock.
For all IoT devices in our lab, we trigger each of their activities $100$ times\footnote{This could be replaced by any reasonably large number.} in order to remove randomness and gain statistically meaningful understanding of the activity packet sequence.
The time interval between two consecutive triggers of the same activity was a random number in the range $[3s, 60s]$.
Here we set the minimum $3$-seconds time interval between two triggers to prevent the smartphone apps from freezing, i.e., becoming unresponsive, due to rapid back-to-back activity triggerings.
The network traffic captured by IoTAthena during these ``active'' period provides the ground truth of activity signatures of IoT device activity.

Filtering the background traffic described in Section~\ref{sec:backgroundnoise}, which happens in parallel with the device activity, leads to an ordered sequence of timestamped IP packets exchanged between IoT devices and the cloud servers.
Each packet in the sequence carries a variety of traffic features such as the timestamp of each packet, local IP address and port number of the IoT device, remote IP address and port number of the cloud server, protocol, packet length, and the actual application payload of IoT applications which are mostly encrypted for security and privacy reasons.
For each packet, we continue to remove features with random and dynamic values due to the protocol designs, e.g., the random local port number at IoT devices in TCP connections with cloud servers and TCP sequence and acknowledge numbers.
In addition, we transform certain traffic features to retain the stable values, e.g., converting dynamic IP addresses of load-balanced cloud servers to the canonical remote cloud server names.

The inter-packet time interval $\tau_j$ between the $j$th packet and the $j+1$th packet in the signature is set to the mean (over the $100$ tries) of the inter-packet time intervals.
To simplify notations, we set $q_1.t$ to $0$, and set $q_{j+1}.t = q_{j}.t + \tau_j$, $j=1, 2, \ldots, n$.

\begin{table*}[hpbt]
\scriptsize
\centering
\caption{Activity signatures of August Lock activities.}
\label{tab:locksig}
\footnotesize
\begin{tabu}{c c |[1pt] c c}
\toprule

\textbf{Activity} & \textbf{Signature} &\textbf{Activity} & \textbf{Signature}\\
\hline

\multirow{12}{*}{\rotatebox{90}{
\begin{tabular}{c}
{\small Opening Mobile App} \\
(12 packets)
\end{tabular}
}} &
\multirow{12}{*}{\tabincell{l}{
0.000s lock:49157 $\rightarrow$ rbs.august.com:443 637B\\
0.132s lock:49157 $\rightarrow$ rbs.august.com:443 221B\\
0.204s rbs.august.com:443 $\rightarrow$ lock:49157 237B\\
0.209s lock:49157 $\rightarrow$ rbs.august.com:443 637B\\
0.327s lock:49157 $\rightarrow$ rbs.august.com:443 237B\\
0.526s rbs.august.com:443 $\rightarrow$ lock:49157 237B\\
0.602s lock:49157 $\rightarrow$ rbs.august.com:443 637B\\
0.723s lock:49157 $\rightarrow$ rbs.august.com:443 237B\\
0.823s rbs.august.com:443 $\rightarrow$ lock:49157 237B\\
1.116s lock:49157 $\rightarrow$ rbs.august.com:443 637B\\
1.205s lock:49157 $\rightarrow$ rbs.august.com:443 221B\\
1.251s rbs.august.com:443 $\rightarrow$ lock:49157 237B
}} & \multirow{21}{*}{\rotatebox{90}{
\begin{tabular}{c}
{\small Manual (Un)Locking} \\
(21 packets)
\end{tabular}
}} & \multirow{21}{*}{\tabincell{l}{
0.000s lock:49157 $\rightarrow$ rbs.august.com:443 637B\\
0.088s lock:49157 $\rightarrow$ rbs.august.com:443 205B\\
0.134s rbs.august.com:443 $\rightarrow$ lock:49157 237B\\
0.441s lock:49157 $\rightarrow$ rbs.august.com:443 637B\\
0.526s lock:49157 $\rightarrow$ rbs.august.com:443 221B\\
0.571s rbs.august.com:443 $\rightarrow$ lock:49157 237B\\
0.581s lock:49157 $\rightarrow$ rbs.august.com:443 637B\\
0.666s lock:49157 $\rightarrow$ rbs.august.com:443 237B\\
0.712s rbs.august.com:443 $\rightarrow$ lock:49157 237B\\
0.870s lock:49157 $\rightarrow$ rbs.august.com:443 637B\\
0.954s lock:49157 $\rightarrow$ rbs.august.com:443 237B\\
1.001s rbs.august.com:443 $\rightarrow$ lock:49157 237B\\
1.078s lock:49157 $\rightarrow$ rbs.august.com:443 637B\\
1.169s lock:49157 $\rightarrow$ rbs.august.com:443 221B\\
1.214s rbs.august.com:443 $\rightarrow$ lock:49157 237B\\
1.321s lock:49157 $\rightarrow$ rbs.august.com:443 637B\\
1.410s lock:49157 $\rightarrow$ rbs.august.com:443 221B\\
1.473s rbs.august.com:443 $\rightarrow$ lock:49157 237B\\
1.559s lock:49157 $\rightarrow$ rbs.august.com:443 637B\\
1.659s lock:49157 $\rightarrow$ rbs.august.com:443 221B\\
1.707s rbs.august.com:443 $\rightarrow$ lock:49157 237B
}}\\
 & & &\\
 & & &\\
 & & & \\
 & & & \\
 & & & \\
 & & & \\
 & & & \\
 & & & \\
 & & & \\
 & & & \\
 & & & \\

\cline{1-2}

\multirow{15}{*}{\rotatebox{90}{
\begin{tabular}{c}
Autolocking \\
(15 packets)
\end{tabular}
}} &
\multirow{15}{*}{\tabincell{l}{
0.000s lock:49157 $\rightarrow$ rbs.august.com:443 637B\\
0.086s lock:49157 $\rightarrow$ rbs.august.com:443 221B\\
0.129s rbs.august.com:443 $\rightarrow$ lock:49157 237B\\
0.240s lock:49157 $\rightarrow$ rbs.august.com:443 637B\\
0.329s lock:49157 $\rightarrow$ rbs.august.com:443 221B\\
0.373s rbs.august.com:443 $\rightarrow$ lock:49157 237B\\
0.990s lock:49157 $\rightarrow$ rbs.august.com:443 637B\\
1.084s lock:49157 $\rightarrow$ rbs.august.com:443 221B\\
1.127s rbs.august.com:443 $\rightarrow$ lock:49157 237B\\
1.277s lock:49157 $\rightarrow$ rbs.august.com:443 637B\\
1.366s lock:49157 $\rightarrow$ rbs.august.com:443 221B\\
1.410s rbs.august.com:443 $\rightarrow$ lock:49157 237B\\
1.549s lock:49157 $\rightarrow$ rbs.august.com:443 637B\\
1.640s lock:49157 $\rightarrow$ rbs.august.com:443 221B\\
1.679s rbs.august.com:443 $\rightarrow$ lock:49157 237B
}} & &\\

 & & & \\
 & & & \\
 & & & \\
 & & & \\
 & & & \\
 & & & \\
 & & & \\
 & & & \\
 \cline{3-4}
 & & \multirow{8}{*}{\rotatebox{90}{
\begin{tabular}{c}
{WiFi (Un)Locking} \\
(8 packets)
\end{tabular}
}} &
\multirow{8}{*}{\tabincell{l}{
0.000s rbs.august.com:443 $\rightarrow$ lock:49156 413B\\
0.008s lock:49156 $\rightarrow$ rbs.august.com:443 605B\\
0.254s rbs.august.com:443 $\rightarrow$ lock:49156 413B\\
0.262s lock:49156 $\rightarrow$ rbs.august.com:443 605B\\
1.426s rbs.august.com:443 $\rightarrow$ lock:49156 413B\\
1.433s lock:49156 $\rightarrow$ rbs.august.com:443 605B\\
1.670s rbs.august.com:443 $\rightarrow$ lock:49156 413B\\
1.678s lock:49156 $\rightarrow$ rbs.august.com:443 605B
}} \\
 & & & \\
 & & & \\
 & & & \\
 & & & \\
 & & & \\

\cline{1-2}

\cline{1-2}

\multirow{6}{*}{\rotatebox{90}{
\begin{tabular}{c}
{Bluetooth} \\
{(Un)Locking}
\end{tabular}
}} &
\multirow{6}{*}{\tabincell{l}{
0.000s  rbs.august.com:443 $\rightarrow$ lock:49156 413B\\
0.016s lock:49156 $\rightarrow$ rbs.august.com:443 605B\\
1.083s rbs.august.com:443 $\rightarrow$ lock:49156 413B\\
1.098s lock:49156 $\rightarrow$ rbs.august.com:443 605B\\
1.379s rbs.august.com:443 $\rightarrow$ lock:49156 413B\\
1.395s lock:49156 $\rightarrow$ rbs.august.com:443 605B
}} & &\\
 & & & \\
 & & & \\
 & & & \\
 & & & \\
 & & & \\

\bottomrule
\end{tabu}
\end{table*}

TABLE~\ref{tab:locksig} illustrates the signatures for various activities of the August Lock\footnote{In our experiments we noticed firmware updates of IoT devices might cause slight changes on the activity signatures.}.
Note that we have used the alternative representation of signatures. 
The signature shown in Fig.~\ref{fig:locksig}(a) corresponds to the lower-right box in TABLE~\ref{tab:locksig}.
The signature shown in Fig.~\ref{fig:locksig}(b) corresponds to the lower-left box in TABLE~\ref{tab:locksig}.
While the sequence of base packets in the signature for {\em Bluetooth (un)locking} (illustrated in Fig.~\ref{fig:locksig}(b)) is a subset of the sequence of base packets in the signature for {\em WiFi (un)locking} (illustrated in Fig.~\ref{fig:locksig}(a)), the additional information embedded in the inter-packet time intervals makes it possible to distinguish these two activities.
With the aid of additional information on inter-packet time intervals, we can distinguish these two activities from network traffic logs, which are difficult to distinguish using the base packets only.

\section{Algorithms for Unveiling IoT Device Activities from Network Traffic}
\label{sec:alg}
\noindent
Having discussed network traffic in Section~\ref{sec:traffic} and device activity signatures in Section~\ref{sec:sig}, we are now ready to present our algorithms for unveiling IoT device activities from network traffic logs.
In Section~\ref{sec:5A}, we formally define the {IoT activity signature matching} problem and the IoT activity extraction problem.
In Section~\ref{sec:5B}, we present the {\tt sigMatch} algorithm for identifying all matches of a given signature in the network traffic log.
In Section~\ref{sec:5C}, we present the {\tt actExtract} algorithm for unveiling the sequence of activities of an IoT device from the network traffic log.
In Section~\ref{sec:5D}, we discuss the limitations and extensions of our algorithms.

\subsection{Problem Formulation}
\label{sec:5A}
\noindent
As discussed in Section~\ref{sec:traffic}, an IoT {\em network traffic log} (denoted by $\mathbb{L}$) is an ordered sequence of packets $(p_1, p_2, \ldots, p_m)$ with increasing timestamps (i.e., $p_{i'}.t < p_{i''}.t$ for $i' < i''$).
As discussed in Section~\ref{sec:sig}, a {\em signature} of an IoT device activity (denoted by $\mathbb{S}$) is an ordered sequence of packets $(q_1, q_2, \ldots, q_n)$ with increasing relative timestamps (i.e., $q_{j'}.t < q_{j''}.t$ for $j' < j''$).
Recall that $\widehat{p}$ and $\widehat{q}$ denote the $7$-tuple obtained by deleting the timestamp in $p$ and the relative timestamp in $q$, respectively.
A {\em signature set} of an IoT device (denoted by $\mathbb{SS}$) is a set of distinct signatures $\{\mathbb{S}^1, \mathbb{S}^2, \ldots, \mathbb{S}^K\}$, one signature per activity of the device.
For ease of presentation, we will use {\it activity} and {\it signature} interchangeably in the rest of the paper.

In light of the end-to-end network latency variations on the Internet~\cite{Dhamdhere:SIGCOMM18,Hoiland:CoNEXT16,Paxson:RFC2988}, we allow an inter-packet time interval tolerance $\epsilon_j > 0$ as the ``safety margin'' for the measurement of $q_{j+1}.t - q_j.t$ when trying to find a match of a signature in the network log.

Let $j$ satisfy $1 < j \le n$ and $\delta > 0$ be a given tolerance.
Let $i'$ and $i^{\prime \prime}$ satisfy $1 \le i' < i^{\prime \prime} \le m$.
We say that $(p_{i'}, p_{i^{\prime \prime}})$ is a {$\delta$-valid match}
of $(q_{j-1}, q_{j})$, if
\begin{enumerate}
\item
$\widehat{p_{i'}} = \widehat{q_{j-1}}$,
$\widehat{p_{i^{\prime \prime}}} = \widehat{q_{j}}$;

\item
\textit{$|({p_{i''}}.t - {p_{i'}}.t) - ({q_j}.t - {q_{j-1}}.t)| \le \delta$.}
%
\end{enumerate}

Let $\mathbb{S} = (q_1, q_2, \ldots, q_n)$ be a signature.
Let $\epsilon = (\epsilon_1, \epsilon_2, \ldots,$ $\epsilon_{n-1})$ be the matching tolerance vector, where $\epsilon_j$ is the tolerance for the matching of $(q_{j}, q_{j+1})$.
Let $(l[1], l[2],  \ldots, l[n])$ be an increasing sequence of integers indicating the index of the location of a packet in the network log.
We say that $(p_{l[1]}, p_{l[2]}, \ldots, p_{l[n]})$ is an $\epsilon$-valid match of signature $\mathbb{S}$ in log $\mathbb{L}$, if $(p_{l[j]}, p_{l[j+1]})$ is an $\epsilon_{j}$-valid match of $(q_{j}, q_{j+1})$, for $j=1, 2, \ldots, n-1$.

We study the following two related problems:

{\bf IoT activity signature matching:}
Given network traffic log $\mathbb{L}$, signature $\mathbb{S}$, and tolerance vector $\epsilon$ for $\mathbb{S}$, identify all $\epsilon$-valid matches of signature $\mathbb{S}$ in log $\mathbb{L}$.

{\bf IoT device activity extraction:}
Given network traffic log $\mathbb{L}$ and signature set $\mathbb{SS}$, find a sequence of IoT activities $\mathbb{A}_1, \mathbb{A}_2, \ldots$, whose execution leads to the network traffic log $\mathbb{L}$.

\subsection{Signature Matching via Time Sensitive Subsequence Matching}
\label{sec:5B}
%
\noindent
The IoT activity signature matching problem is different from the traditional subsequence matching problem~\cite{Maier:ACM78} and the longest common subsequence problem~\cite{Bergroth:SPIRE00, Cormen:book09} due to the inter-packet time interval constraint. 
The matching problem with such constraints cannot be solved via the simple adjustment of existing algorithms.
We solve the signature matching problem using a time-sensitive subsequence matching approach, called {\tt sigMatch}, as presented in Algorithm~\ref{alg-01}.

\begin{algorithm}[htbp]
\caption{${\rm \bf sigMatch}(\mathbb{L}, \mathbb{S}, \epsilon)$}
\label{alg-01}
\LinesNumbered
\KwIn{Network traffic log $\mathbb{L} = (p_{1}, p_{2}, \ldots, p_{m})$,
Signature $\mathbb{S} = (q_{1}, q_{2}, \ldots, q_{n})$,
tolerance vector $\epsilon =(\epsilon_1, \epsilon_2, \ldots, \epsilon_{n-1})$.
}
\KwOut{A DAG $G_{\mathbb{LS}} = (V_{\mathbb{LS}}, E_{\mathbb{LS}})$ that
captures all $\epsilon$-valid matches of signature $\mathbb{S}$ in
$\mathbb{L}$.}
$V_\mathbb{LS} \leftarrow \emptyset$;
$E_\mathbb{LS} \leftarrow \emptyset$;

\For {$i:=1$ \KwTo $m$}{
    \If { \textit{$\widehat{p_i} == \widehat{q_1}$} }{
        $V_\mathbb{LS} \leftarrow V_\mathbb{LS} \cup \{v_{i, 1}\}$;
    }
    \For {$j:=2$ \KwTo $n$}{
        \For {$k:=1$ \KwTo $i-1$}{
            \If {\textit{
            $v_{k, j-1} \in V_\mathbb{LS}$
            {\bf and}
            $(p_k, p_i)$ is an $\epsilon_{j-1}$-valid match of $(q_{j-1}, q_j)$
            }}{
              $V_\mathbb{LS} \leftarrow V_\mathbb{LS} \cup \{v_{i, j}\}$;

              $E_\mathbb{LS} \leftarrow E_\mathbb{LS} \cup \{(v_{i, j}, v_{k, j-1})\}$;
            }
        }
    }
}
{\bf output} DAG $G_\mathbb{LS}$.
\end{algorithm}

For a given network traffic log $\mathbb{L} = (p_1, p_2, \ldots, p_m)$ and signature $\mathbb{S} = (q_1, q_2, \ldots, q_n)$, together with a inter-packet time interval tolerance vector $\epsilon$, we compute a DAG $G_\mathbb{LS} = (V_\mathbb{LS}, E_\mathbb{LS})$ that captures all $\epsilon$-valid matches of signature $\mathbb{S}$ in log $\mathbb{L}$.
The vertex set $V_\mathbb{LS}$ contains vertices in the form of $v_{i, j}$, where $p_i$ is a {\em potential match} of $q_j$.
The edge set $E_\mathbb{LS}$ contains directed edges in the form of $(v_{i, j}, v_{k, j-1})$, where $(p_k, p_i)$ is an $\epsilon_{j-1}$-valid match of $(q_{j-1}, q_j)$ for $ 1 \le k \le i-1$, and there is a directed path from vertex $v_{i, j}$ to a vertex $v_{i', 1} \in V_\mathbb{LS}$ (for some $i' \le i-j+1$).

If $\widehat{p_i} \ne \widehat{q_j}$, vertex $v_{i,j}$ {\em does not} exist.
If $\widehat{p_i} = \widehat{q_j}$, vertex $v_{i,j}$ {\em may} exist.
Each edge has the form $(v_{i, j}, v_{k, j-1})$ for some $k < i$.
Hence we have $|V_\mathbb{LS}| \le mn$ and $|E_\mathbb{LS}| \le \frac{m (m-1) (n-1)}{2}$.

In Line 1 of Algorithm~\ref{alg-01}, both $V_{\mathbb{LS}}$ and $E_{\mathbb{LS}}$ are initialized to $\emptyset$.
The algorithm then populates the vertex set and the edge set while looping over the packets $p_1, p_2, \ldots, p_m$.
For each $i$, the algorithm loops over the packets $q_1, q_2, \ldots, q_n$.
When $\widehat{p_i} = \widehat{q_1}$, $v_{i, 1}$ is a vertex in the DAG.
For $j=2, 3, \ldots, n$, $v_{i, j}$ is a vertex if and only if $\widehat{p_i} = \widehat{q_j}$ {\bf and} $(p_k, p_i)$ is an $\epsilon_{j-1}$-valid match of $(q_{j-1}, q_j)$ for some $k < i$.
In this case, $(v_{i, j}, v_{k, j-1})$ is an edge in the DAG.

\begin{figure}[htbp]
\centering
\includegraphics[width=3.2in]{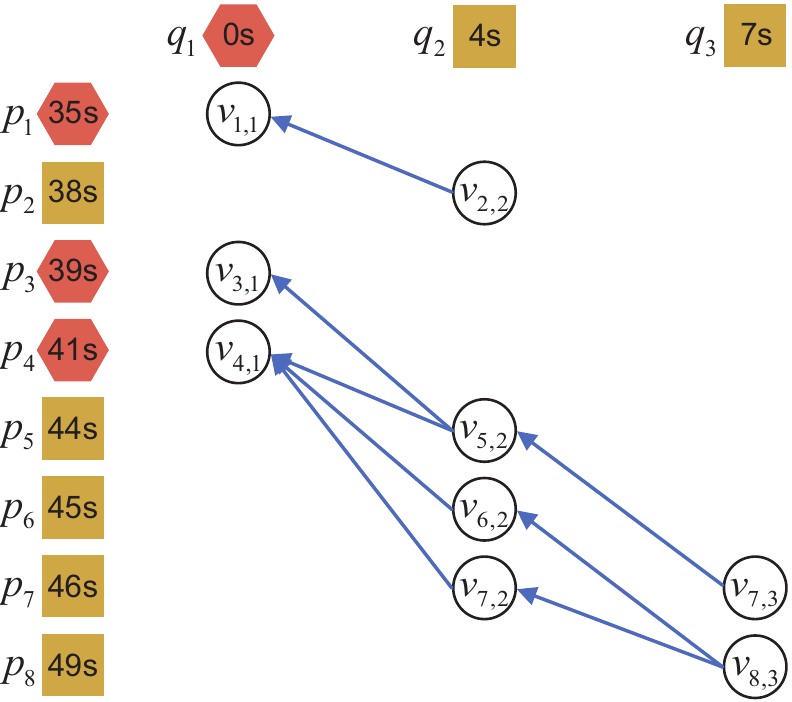}
\caption{Running example of sigMatch: row index corresponds to the traffic log, column index corresponds to the signature.}
\label{fig:match}
\vspace{-0.15in}
\end{figure}

We use Fig.~\ref{fig:match} to illustrate a running example of {\tt sigMatch}.
The goal is to identify all $\epsilon$-valid matches of signature $\mathbb{S} = (q_1, q_2, q_3)$ in log $\mathbb{L} = (p_1, p_2, p_3, p_4, p_5, p_6, p_7, p_8)$ with tolerance vector $\epsilon=(1, 1)$.
In this example, we have $\widehat{p_1} = \widehat{p_3} = \widehat{p_4} = \widehat{q_1}$, denoted by the {\em hexagon} shape; we also have $\widehat{p_2} = \widehat{p_5} = \widehat{p_6} = \widehat{p_7}= \widehat{p_8}= \widehat{q_2}= \widehat{p_3}$, denoted by the {\em square} shape.
The timestamps (for traffic log) and relative timestamps (for signature) are inside the corresponding shape.

We start from $p_1$.
Since $\widehat{p_1} = \widehat{q_1}$, vertex $v_{1,1}$ is added to $V_\mathbb{LS}$;
Then we move to $p_2$.
Since $\widehat{p_2} = \widehat{q_2}$, $v_{1,1} \in V_{LS}$, and $|({p_2}.t - {p_1}.t) - ({q_2}.t - {q_1}.t)| = |(38 - 35) - (4 - 0)| \le 1$, vertex $v_{2,2}$ is added to $V_\mathbb{LS}$ and directed edge $(v_{2,2}, v_{1,1})$ is added to $E_\mathbb{LS}$.
Similarly, vertices $v_{3,1}$ and $v_{4,1}$ are added to $V_\mathbb{LS}$.
Next, we pay attention to the row corresponding to $p_5$.
We found $\widehat{p_5} = \widehat{q_2}$.
For $k=1$, we found $v_{1, 1} \in V_\mathbb{LS}$, but the time interval does not match.
For $k=3$, we found $v_{3, 1} \in V_\mathbb{LS}$, and the time interval matches.
Hence vertex $v_{5, 2}$ is added to $V_\mathbb{LS}$, and edge $(v_{5, 2}, v_{3, 1})$ is added to $E_\mathbb{LS}$.
For $k=4$, we found $v_{4, 1} \in V_\mathbb{LS}$, and the time interval matches.
At this moment, vertex $v_{5, 2}$ is already in $V_\mathbb{LS}$, and edge $(v_{5, 2}, v_{4, 1})$ is added to $E_\mathbb{LS}$.
We obtain the DAG as shown in Fig.~\ref{fig:match} by continuing the above process.

\begin{theorem}
Algorithm {\tt sigMatch} has a worst-case time complexity of $O(m^2 n)$, where $n$ is the number of packets in the signature $\mathbb{S}$, and $m$ is the number of packets in the network traffic log $\mathbb{L}$.
Furthermore,
\begin{enumerate}
\item [(a)]
If $(p_{l[1]}, p_{l[2]}, \ldots, p_{l[n]})$ is an $\epsilon$-valid match of $\mathbb{S}$ in $\mathbb{L}$, then $(v_{l[n], n}, v_{l[n-1], n-1}, \ldots, v_{l[1], 1})$ is a directed path in $G_\mathbb{LS}$, and $l[1] < l[2] < \cdots < l[n]$.

\item [(b)]
If $(v_{l[n], n}, v_{l[n-1], n-1}, \ldots, v_{l[1], 1})$ is a directed path in $G_\mathbb{LS}$, then $(p_{l[1]}, p_{l[2]}, \ldots, p_{l[n]})$ is an $\epsilon$-valid match of $\mathbb{S}$ in $\mathbb{L}$, and $l[1] < l[2] < \cdots < l[n]$.
\end{enumerate}
\end{theorem}
{\bf Proof.}
The loop over $i$ runs $m$ times.
The loop over $j$ runs $n$ times.
The loop over $k$ runs $i-1$ times, for each $i$.
This leads to the worst-case time complexity of $O(m^2 n)$.

From the condition in Line 7 of the algorithm, we notice that {\em there is an edge in the form $(v_{i, j}, v_{k, j-1})$ {\bf if and only if} there is an $\epsilon_{[j]}$-valid match of $(q_1, q_2, \ldots, q_j)$ in $\mathbb{L}$ that matches $(q_{j-1}, q_j)$ to $(p_k, p_i)$}, where $\epsilon_{[j]} = (\epsilon_1, \epsilon_2, \ldots, \epsilon_{j-1})$.
This leads to claims (a) and (b).
\hfill$\Box$

We point out that the total number of $\epsilon$-valid matches of signature $\mathbb{S}$ in $\mathbb{L}$ may be exponential.
However, all of them are captured by a polynomial sized DAG $G_\mathbb{LS}$, which can be computed in polynomial time.

\subsection{Unveiling IoT Activities from Network Traffic Log}
\label{sec:5C}
\noindent
We investigate how to unveil the activities of an IoT device using {\tt sigMatch} in Algorithm~\ref{alg-01} as a building block.
Note that we can separate the traffic of a specific IoT device from all network traffic using the IoT device's distinct IP address.
For a given IoT device, we first extract its signature set $\mathbb{SS} = \{\mathbb{S}^1, \mathbb{S}^2, \ldots, \mathbb{S}^K\}$.
For each signature $\mathbb{S}^k$, using its corresponding tolerance vector $\epsilon^k$, we can apply {\tt sigMatch} to construct the corresponding DAG $G_{\mathbb{LS}^k}$ in $O(m^2 n_k)$ worst-case time, where $n_k$ is the number of packets in $\mathbb{S}^k$.
We can compute all $K$ DAGs in $O(K m^2 n_{\max})$ worst-case time, where $n_{\max} = \max \{n_1, n_2, \ldots, n_K\}$.

For each $k=1, 2, \ldots, K$, there may be zero or more $\epsilon^k$-valid matches of signature $\mathbb{S}^k$.
Making use of $G_{\mathbb{LS}^k}$, we can either confirm that there is no $\epsilon^k$-valid match (when there is no vertex $v_{i, n_k}$ in $V_{\mathbb{LS}^k}$) or compute {\em the earliest} $\epsilon^k$-valid match $(p_{l[1]}, p_{l[2]}, \ldots, p_{l[n_k]})$, in the sense that $(p_{l[1]}, p_{l[2]}, \ldots, p_{l[n_k]})$ is {\em lexicographically smallest}, in $O(m + n_k)$ worst-case time.

Given the network traffic $\mathbb{L}$, and the valid matches of signatures in $\mathbb{SS}$, how do we decide which IoT activity happened first?
Through extensive experiments, we found that {\em in normal situations, each network packet corresponding to an earlier IoT activity proceeds every network packet corresponding to a later IoT activity}.
Therefore the signature that has the earliest match happens first.
Once this decision is made, we can delete each packet with a timestamp no later than that of the last packet in the match of the found signature from the network traffic.
Repeating the above process, we can unveil the sequence of IoT activities from the given network traffic.
We formally describe this process called {\tt actExtract} in Algorithm~\ref{alg-02}.

\begin{algorithm}[htbp]
\caption{${\rm \bf actExtract}(\mathbb{L}, \mathbb{SS},  \varepsilon)$
}
\label{alg-02}
\LinesNumbered
\KwIn{
Network traffic
$\mathbb{L} = (p_{1}, p_{2}, \ldots, p_{m})$,
signature set
$\mathbb{SS} = \{\mathbb{S}^{1}, \mathbb{S}^{2}, \ldots, \mathbb{S}^{K}\}$,
$\varepsilon = (\epsilon^1, \epsilon^2, \ldots, \epsilon^K)$
where $\epsilon^k$ is the match tolerance vector for
$\mathbb{S}^k$.}
\KwOut{A sequence of IoT activities $\mathbb{A}_1, \mathbb{A}_2, \ldots$.}

\For {$k:=1$ \KwTo $K$}{
    $G_{\mathbb{S}^k} \leftarrow sigMatch(\mathbb{L}, \mathbb{S}^k, \epsilon^k)$;
}

\While {\textit{some signature $\mathbb{S}^k$ has a match in $G_{\mathbb{S}^k}$}}{
Let $\mathbb{S}^{k'}$ have the earliest match;

{\bf output} Activity corresponding to $\mathbb{S}^{k'}$;

Remove all packets in $\mathbb{L}$ with timestamp no later than that of the
    last matched packet for $\mathbb{S}^{k'}$.

\For {$k:=1$ \KwTo $K$}{
    $G_{\mathbb{S}^k} \leftarrow sigMatch(\mathbb{L}, \mathbb{S}^k, \epsilon^k)$;
}
}

\end{algorithm}
\begin{theorem}
The worst-case time complexity of Algorithm~\ref{alg-02} is $O(K m^3 n_{\max})$, where $K$ is the number of signatures, $n_{\max}$ is the maximum number of packets in any of the signatures, and $m$ is the number of packets in network traffic log $\mathbb{L}$.
In normal situations (i.e., each packet for an earlier activity precedes every network packet of a later activity), {\tt actExtract} correctly outputs a sequence of IoT activities $\mathbb{A}_1, \mathbb{A}_2, \ldots$ whose sequential execution will generate a network traffic log that may be different from $\mathbb{L}$ only in the timestamp fields.
\end{theorem}
{\bf Proof.}
Initially, the $K$ DAGs can be computed in $O(K m^2 n_{\max})$ time.
The earliest match of $\mathbb{S}^k$ can be computed in $O(m + n_k)$ time, $\forall k$.
Selecting the signature with the earliest match requires $O(K n_{\max})$ time.
This process is repeated for no more than $m$ times, hence the time complexity.

Next, we prove the correctness of the algorithm.
Assuming that the sequence of IoT device activities that generated the network traffic $\mathbb{L}$ is $\mathbb{A}_1, \mathbb{A}_2, \ldots, \mathbb{A}_x$.
By our {\em normal} assumption, each network packet of $\mathbb{A}_1$ must happen earlier than every network packet of $\mathbb{A}_\lambda$, for any $\lambda > 1$.
Since {\tt actExtract} uses the earliest match, it will output $\mathbb{A}_1$ as the first activity, and all of the packets in the computed match for $\mathbb{A}_1$ have timestamps earlier than the timestamp of any packet in other IoT activity $\mathbb{A}_\lambda$, with $\lambda > 1$.
Hence, when we delete the packets matched for $\mathbb{A}_1$, we delete all of the packets generated for $\mathbb{A}_1$, but none of the packets generated by $\mathbb{A}_\lambda$ with $\lambda > 1$.
Therefore {\tt actExtract} will next output $\mathbb{A}_2$, then $\mathbb{A}_3$, and so on.
This proves the correctness of the algorithm.
\hfill
$\Box$

When we execute the computed sequence of IoT device activities, the network traffic observed may be different from $\mathbb{L}$, but only in the timestamp field.
The sequence of packets will have increasing timestamps.
Ignoring the timestamp field, two sequences of packets will be identical with $\mathbb{L}$.
Note that we can divide the network traffic log into multiple sublogs where each sublog corresponds to a unique IoT device.
We can apply {\tt actExtract} to each sublog in parallel to {\em unveil the IoT activities for all IoT devices}.

\subsection{Discussions}
\label{sec:5D}
\noindent
Our proposed {\tt actExtract} algorithm can unveil the activity sequence of an IoT device with no ambiguity and guarantee correctness, assuming there is no ongoing attack and the device can only carry out one activity at a time, which is true for most devices.
For devices that allow two or more concurrent activities, such as IP cameras, we can modify {\tt sigMatch} algorithm to record only non-overlapping matches in network traffic for a signature in a DAG.
We can then build the DAG for the same device's signatures independently and output all the identified activities.
In case when there is attacking traffic, it is possible that one packet is matched to two different signatures. We can add a variable in {\tt sigMatch} to record all the signatures that a packet is matched to. If such a conflict happens, our algorithm can report it as an anomaly and raise an alarm.

\section{Experimental Evaluations}
\label{sec:eval}
\noindent
We evaluate the performance of IoTAthena using two different settings:
1) our own smart home testbed,
and
2) a large public IoT network traffic dataset~\cite{Ren:IMC19}.
We first describe these settings in Section~\ref{sec:evalA}.
In Section~\ref{sec:evalB}, we present experimental results on the sensitivity of IoTAthena's accuracy on the matching tolerance.
In Section~\ref{sec:evalC}, we present experimental results for homogeneous device activities.
In Section~\ref{sec:evalD}, we present experimental results for mixed device activities, together with a case study.

\subsection{Experiment Setting}
\label{sec:evalA}
\noindent
Our smart home testbed has $16$ widely-used IoT devices, including multiple models of {\em IP cameras, smart bulbs, smart doorbells, smart locks, and smart plugs}.
These IoT devices are all ranked as popular by Smart Home DB~\cite{smarthomedb:web}. 
Our experiments have identified a total of 44 different device activities by using these 16 devices.
The numbers of devices and activities in this study are comparable to existing studies on understanding IoT device activities in smart home network environments~\cite{Acar:WiSec2020,Trimananda:NDSS20}. 
This controlled smart home environment was used to create the ``silent" week for collecting, analyzing, and characterizing background network traffic, as discussed in Section~\ref{sec:traffic}.
For each IoT device activity, we repeatedly generate the activity while collecting the associated network traffic as well as recording the activity logs which are used for establishing the ground truth at the same time.
Using the signature extraction technique introduced in Section~\ref{sec:sig}, we were able to extract signatures for most of the device activities of all $16$ representative IoT devices except the {\em stream off} activity of Amcrest ProHD camera, which does not have the deterministic traffic pattern to form a signature.

In addition to network traffic and activity logs collected from our own smart home testbed, we also evaluated IoTAthena's performance using a large public IoT network traffic dataset~\cite{Ren:IMC19}, known as the MON(IOT)R dataset.
The MON(IOT)R dataset includes raw IP data traffic and the labeled activity logs of $25$ IoT devices\footnote{We evaluated IoTAthena on the IoT device activities with at least $30$ samples in the MON(IOT)R dataset in order to have statistically meaningful results.}.
Among these devices, $6$ of them are also included in our smart home testbed, while the other $19$ devices are unique to the dataset.
The IoT network traffic and labeled activities in the dataset allow us to evaluate the performance of IoTAthena.

\subsection{Sensitivity Analysis on the Tolerance Parameter}
\label{sec:evalB}
\noindent
Our time-sensitive subsequence matching algorithm {\tt sigMatch} uses the tolerance vector $\epsilon = (\epsilon_1, \epsilon_2, \ldots, \epsilon_{n-1})$ for accommodating inter-packet time intervals' variations.
In our experiment, $\epsilon_j$ is set to $r \times \sigma_j$ for $j \in [1, n-1]$, where $\sigma_j$ is the standard deviation of the inter-packet time interval between two consecutive packets $q_{j}$ and $q_{j+1}$ and $r \ge 1$ is a tunable parameter.

\begin{figure*}[htbp]
\centering
\subfigure[{\scriptsize TP-Link Bulb on}]{
\includegraphics[width=1.45in]{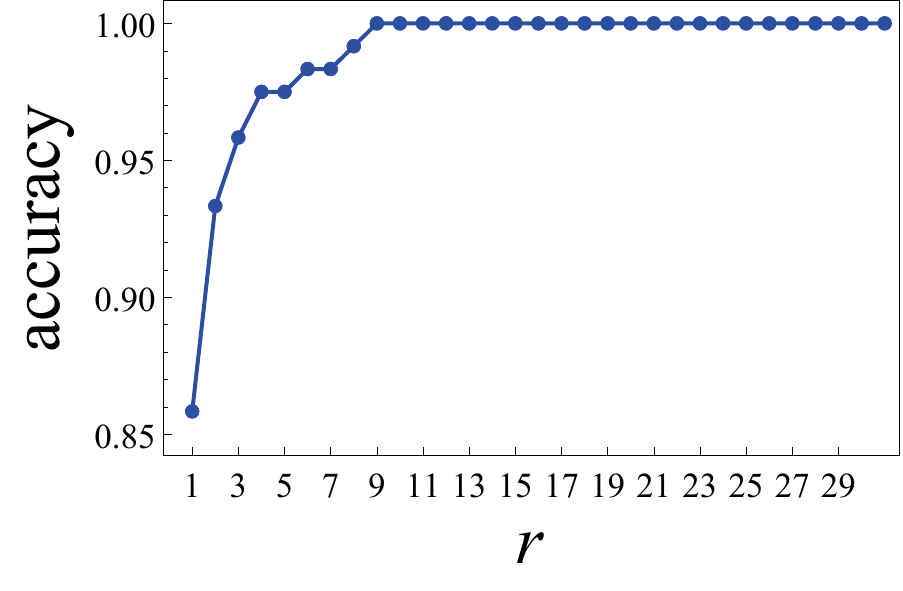}
}
\subfigure[{\scriptsize TP-Link Bulb off}]{
\includegraphics[width=1.45in]{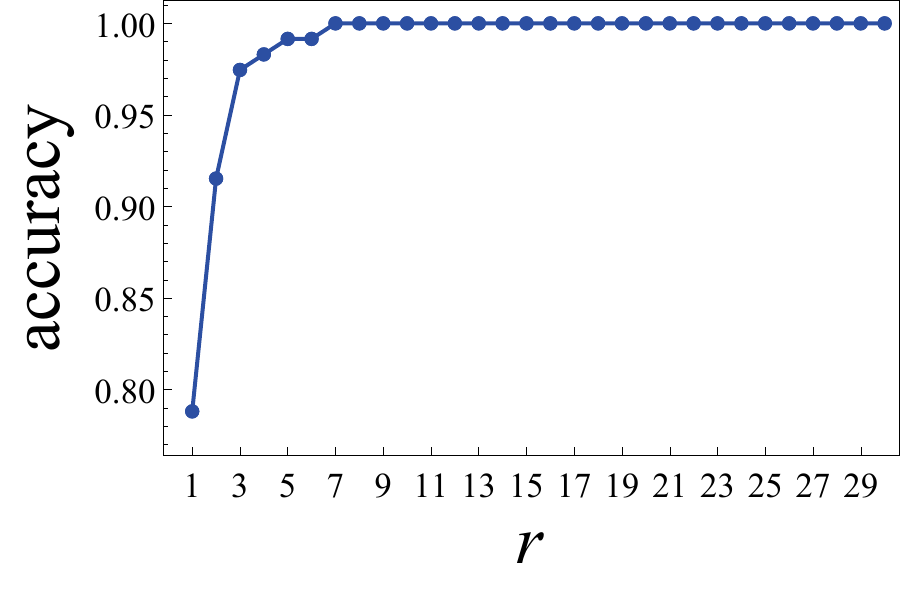}
}
\subfigure[{\scriptsize TP-Link Plug on}]{
\includegraphics[width=1.45in]{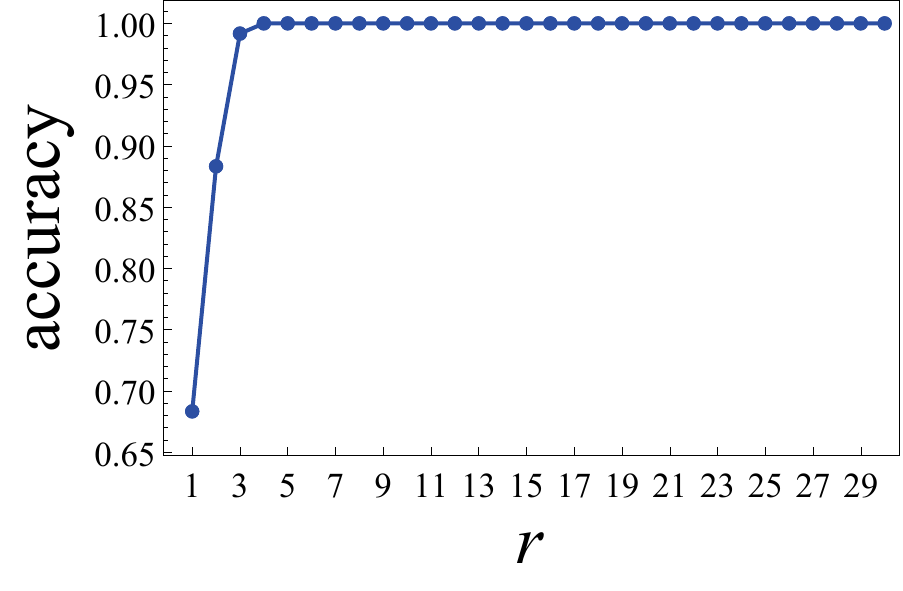}
}
\subfigure[{\scriptsize TP-Link Plug off}]{
\includegraphics[width=1.45in]{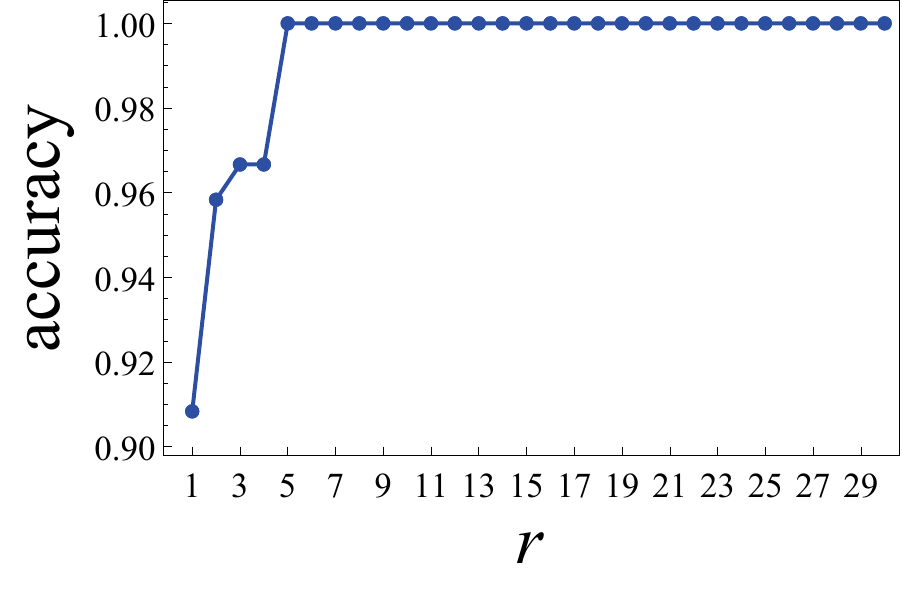}
}

\subfigure[{\scriptsize August Lock app opening}]{
\includegraphics[width=1.45in]{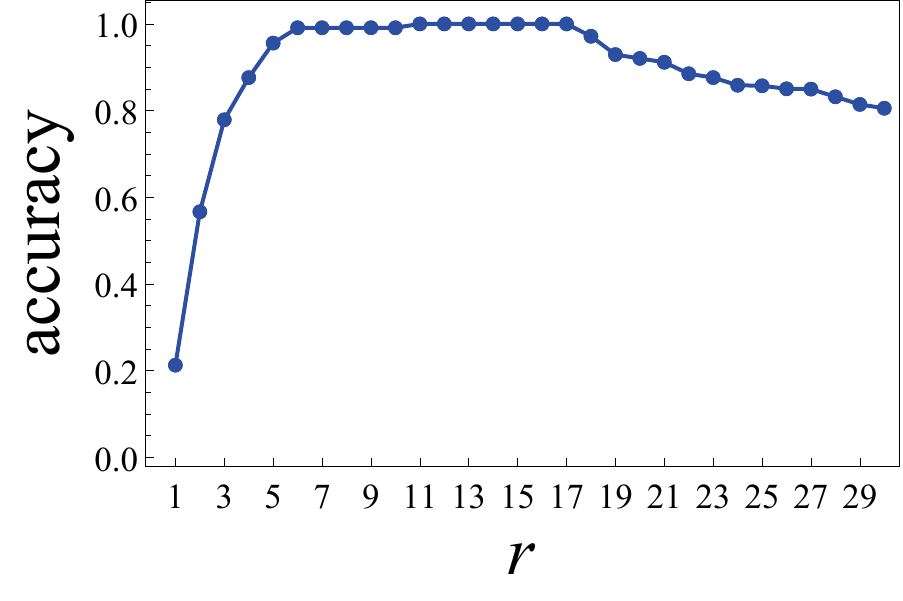}
}
\subfigure[{\scriptsize August Lock WiFi (un)locking}]{
\includegraphics[width=1.45in]{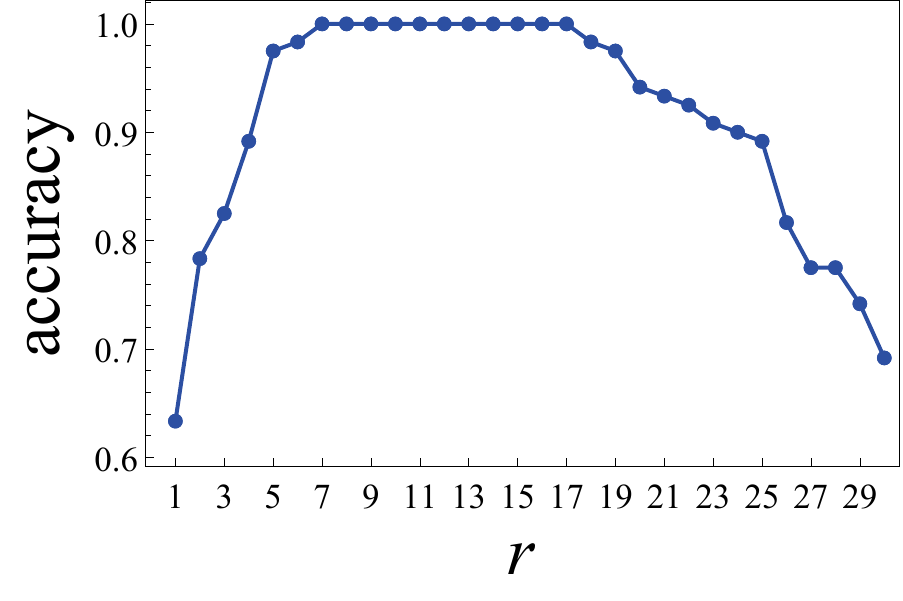}
}
\subfigure[{\scriptsize August Lock autolocking}]{
\includegraphics[width=1.45in]{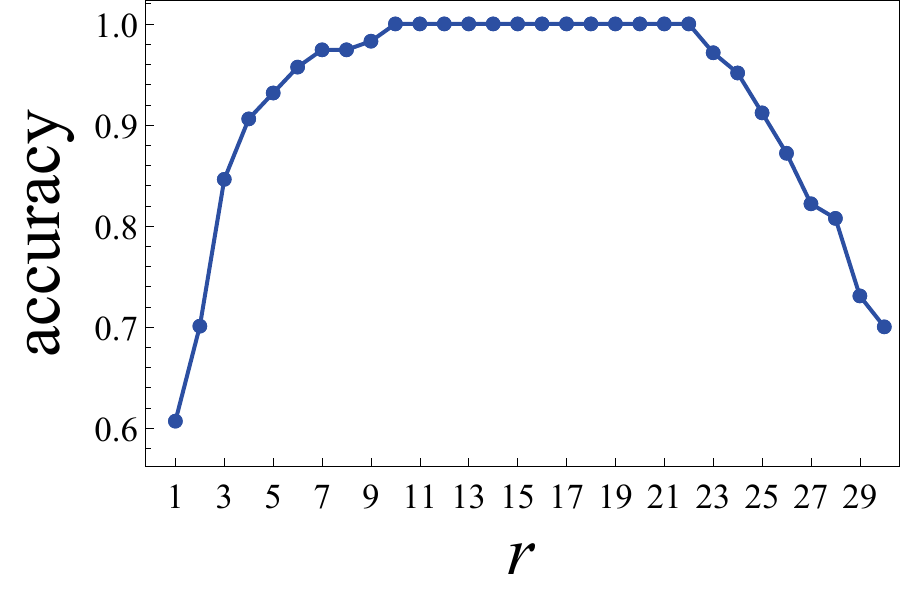}
}
\subfigure[{\scriptsize August Lock BLE (un)locking}]{
\includegraphics[width=1.45in]{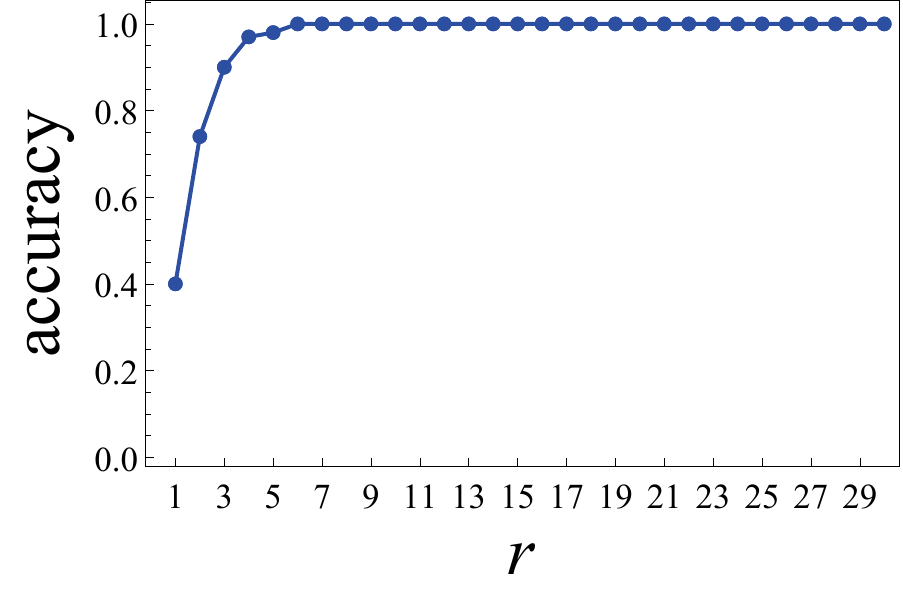}
}
\caption{The impact of $r$ in the inter-packet time interval tolerance
parameter on the accuracy of IoT activity signature matching.}
\label{fig:sensitivity}
\end{figure*}

The accuracy of IoTAthena depends on the matching tolerance parameter.
Intuitively, when the matching tolerance is very small, IoTAthena tends to unveil fewer activities due to the strict checking of inter-packet time intervals, leading to low accuracy. 
On the other hand, when the matching tolerance is very large, IoTAthena tends to have more false negatives due to the loose checking of inter-packet time intervals, also leading to low accuracy.
In order to have a deeper understanding of this dependency, we carried out sensitivity analysis.
For each device activity, we repeatedly triggered it $120$ times with random delays between two consecutive experiments.
We then ran $6$-fold cross validation using the data collected.
In each of the $6$ rounds, we observe the accuracy of IoTAthena on $20$ of the experiments, while using the remaining $100$ to generate the signature.
Fig.~\ref{fig:sensitivity} illustrates some representative results.

Figs.~\ref{fig:sensitivity}(a)-(d) show the accuracy changes for unveiling the {\it on} and {\it off} activities of TP-Link Bulb and TP-Link Plug as $r$ increases from $1$ to $30$, while Figs.~\ref{fig:sensitivity}(e)-(h) show the accuracy dynamics of unveiling August Lock's app opening, Wifi (un)locking, autolocking, and Bluetooth (un)locking activities.
In Figs.~\ref{fig:sensitivity}(a)-(d), we observe that the accuracy of IoTAthena exhibits a non-decreasing trend for the {\it on} and {\it off} activities of both TP-Link Bulb and TP-Link Plug, when $r$ increases from $1$ to $30$.
These observations are not surprising since the increasing value of $r$ leads to a higher tolerance value to allow larger inter-packet time intervals.

However, Figs.~\ref{fig:sensitivity}(e)-(g) contradict such conjectures as
increasing $r$ to a particular value leads to decreasing accuracy in matching
August Lock's app opening, WiFi (un)locking, and autolocking activities.
Our in-depth investigation discovered that the accuracy decrease for the
larger $r$ values is due to the interference of August Lock's background
traffic noise. The background traffic happens to shares overlapping 
packets with the signatures of app opening, WiFi (un)locking, and 
autolocking activities when we allow bigger tolerance
of inter-packet intervals.
Unlike Figs.~\ref{fig:sensitivity}(e)-(g), the accuracy of unveiling
August Lock's Bluetooth (un)locking activities changes in a non-decreasing
fashion in Fig.~\ref{fig:sensitivity}(h) as $r$ increases from $1$ to $30$.
The underlying reason of this distinct observation is the unique traffic
patterns of the network traffic collected when triggering August Lock's
Bluetooth (un)locking activities, where no background noise that cannot
be filtered out has been observed.

In summary, our sensitivity analysis on the tolerance parameter confirmed
the importance and impact of choosing appropriate tolerance values during
the IoT device activity extraction process.
More importantly, the observations in Fig.~\ref{fig:sensitivity} highlight
the rationale and necessity of our {\it full} packet sequences with
inter-packet time intervals as the IoT device activity signature and our
time-sensitive subsequence matching algorithm for unveiling IoT device activities.

\subsection{Performance of IoTAthena on Homogeneous Device Activities}
\label{sec:evalC}
\noindent
Having done the sensitivity analysis, we focused on
evaluating IoTAthena's performance of unveiling homogeneous device activities.
We first present experimental results on our smart home
testbed.
We then present results on the MON(IOT)R dataset~\cite{Ren:IMC19}.

For each IoT device activity, we repeated it $120$ times
and collected the network traffic on the router in
our smart home testbed.
We again ran the $6$-fold cross validation.
TABLE~\ref{tab:evaluate} shows the
accuracy ($\mathcal{A}$),
precision ($\mathcal{P}$),
and recall ($\mathcal{R}$)
measures
of IoTAthena for various activities of the $16$ devices in our
smart home testbed, with $r$ set to $3$, $11$, and $23$,
respectively.

\begin{table*}[htbp]
\scriptsize
\centering
\caption{Accuracy, precision, and recall of IoTAthena for $16$ devices with $r$ set as $3$, $11$, $23$ respectively.}
\vspace{-0.15in}
\label{tab:evaluate}
\begin{tabular}{l l l  l l l | l l l | l l l}
\toprule
\multirow{2}{*}{\textbf{Type}} & \multirow{2}{*}{\textbf{Device Name}} &\multirow{2}{*}{\textbf{Activity}} & \multicolumn{3}{c|}{\textbf{$r=3$}} & \multicolumn{3}{c|}{\textbf{$r=11$}} & \multicolumn{3}{c}{\textbf{$r=23$}} \\
 &  &  &  \textbf{$\mathcal{A}$} & \textbf{$\mathcal{P}$} & \textbf{$\mathcal{R}$}
 & \textbf{$\mathcal{A}$} & \textbf{$\mathcal{P}$} & \textbf{$\mathcal{R}$}
 & \textbf{$\mathcal{A}$} & \textbf{$\mathcal{P}$} & \textbf{$\mathcal{R}$}\\\hline
\multirow{9}{*}{bulb}  &  \multirow{2}{*}{Philips  Hue} &  on or off & $0.98$ & $1.00$ & $0.98$ & $1.00$ & $1.00$ & $1.00$ & $1.00$ & $1.00$ & $1.00$ \\
& & brightness & $0.95$ & $1.00$ & $0.95$ & $1.00$ & $1.00$ & $1.00$ & $1.00$ & $1.00$ & $1.00$ \\
\cline{2-12}
  & \multirow{3}{*}{Sengled  SmartLED} & on & $0.88$ & $1.00$ & $0.88$ & $1.00$ & $1.00$ & $1.00$ & $1.00$ & $1.00$ & $1.00$\\
  &  & off & $0.89$ & $1.00$ & $0.89$ & $1.00$ & $1.00$ & $1.00$ & $1.00$ & $1.00$ & $1.00$\\
 &  & brightness & $0.92$ & $1.00$ & $0.92$ & $1.00$ & $1.00$ & $1.00$ & $1.00$ & $1.00$ & $1.00$\\
\cline{2-12}
  & \multirow{4}{*}{TP-Link Bulb}  & on & $0.96$ & $1.00$ & $0.96$ & $1.00$ & $1.00$ & $1.00$ & $1.00$ & $1.00$ & $1.00$\\
  &   & off & $0.97$ & $1.0$0 & $0.97$ & $1.00$ & $1.00$ & $1.00$ & $1.00$ & $1.00$ & $1.00$\\
  &   & color & $0.99$ & $1.00$ & $0.99$ & $1.00$ & $1.00$ & $1.00$ & $1.00$ & $1.00$ & $1.00$\\
  &   & brightness & $0.95$ & $1.00$ & $0.95$ & $1.00$ & $1.00$ & $1.00$ & $1.00$ & $1.00$ & $1.00$\\
\cline{1-12}
\multirow{13}{*}{camera}
 &  Amcrest ProHD & stream on &  $0.86$ &  $1.00$ &  $0.86$ &  $1.00$ &  $1.00$ &  $1.00$ & $1.00$ &  $1.00$ &  $1.00$\\
\cline{2-12}
 & \multirow{3}{*}{Arlo - Q Indoor} & stream on & $0.96$ & $1.00$ & $0.96$ & $1.00$ & $1.00$ & $1.00$ & $1.00$ & $1.00$ & $1.00$\\
 &  & stream off & $0.94$ & $1.00$ & $0.94$ & $1.00$ & $1.00$ & $1.00$ & $1.00$ & $1.00$ & $1.00$\\
 &  & motion detection & $0.98$ & $1.00$ & $0.98$ & $1.00$ & $1.00$ & $1.00$ & $1.00$ & $1.00$ & $1.00$\\
\cline{2-12}
  & \multirow{3}{*}{Arlo Ultra} & stream on &  $0.88$ &  $1.00$ &  $0.88$ &  $1.00$ &  $1.00$ &  $1.00$ &  $1.00$ &  $1.00$ &  $1.00$\\
  &  & stream off &  $0.92$ &  $1.00$ &  $0.92$ &  $1.00$ &  $1.00$ &  $1.00$ &  $1.00$ &  $1.00$ &  $1.00$\\
  &  & motion detection & $0.95$ & $1.00$ & $0.95$ & $1.00$ & $1.00$ & $1.00$ & $1.00$ & $1.00$ & $1.00$\\
\cline{2-12}
  & \multirow{3}{*}{Blink XT2} & stream on & $0.92$ & $1.00$ & $0.92$ & $0.99$ & $1.00$ & $0.99$ & $0.99$ & $1.00$ & $0.99$\\
  &  & stream off &  $0.95$ &  $1.00$ &  $0.95$ &  $1.00$ &  $1.00$ &  $1.00$ &  $1.00$ &  $1.00$ &  $1.00$\\
  &  & motion detection & $0.96$ & $1.00$ & $0.96$ & $1.00$ & $1.00$ & $1.00$ & $1.00$ & $1.00$ & $1.00$\\
\cline{2-12}
  & \multirow{3}{*}{Reolink Camera} & stream on &  $0.98$ &  $1.00$ &  $0.98$ &  $1.00$ &  $1.00$ &  $1.00$ &  $1.00$ &  $1.00$ &  $1.00$ \\
  &  & stream off &  $0.96$ &  $1.00$ &  $0.96$ &  $1.00$ &  $1.00$ &  $1.00$ &  $1.00$ &  $1.00$ &  $1.00$\\
  &  & motion detection & $1.00$ & $1.00$ & $1.00$ & $1.00$ & $1.00$ & $1.00$ & $1.00$ & $1.00$ & $1.00$\\
\cline{1-12}
\multirow{8}{*}{doorbell}  &  \multirow{4}{*}{August Doorbell Cam Pro} & stream on & $1.00$ & $1.00$ & $1.00$ & $1.00$ & $1.00$ & $1.00$ & $1.00$ & $1.00$ & $1.00$\\
 &  & stream off &  $1.00$ &  $1.00$ &  $1.00$ &  $1.00$ &  $1.00$ &  $1.00$ &  $1.00$ &  $1.00$ &  $1.00$\\
 &  & ringing & $0.94$ & $1.00$ & $0.94$ & $1.00$ & $1.00$ & $1.00$ & $1.00$ & $1.00$ & $1.00$\\
 &  & motion detection & $0.95$ & $1.00$ & $0.95$ & $1.00$ & $1.00$ & $1.00$ & $1.00$ & $1.00$ & $1.00$\\
\cline{2-12}
 &  \multirow{4}{*}{Ring VideoDoorbell} & stream on & $0.98$ & $1.00$ & $0.98$ & $1.00$ & $1.00$ & $1.00$ & $1.00$ & $1.00$ & $1.00$\\
 & & stream off &  $0.96$ &  $1.00$ &  $0.96$ &  $1.00$ &  $1.00$ &  $1.00$ &  $1.00$ &  $1.00$ &  $1.00$\\
 &  & ringing & $0.96$ & $1.00$ & $0.96$ & $1.00$ & $1.00$ & $1.00$ & $1.00$ & $1.00$ & $1.00$\\
  &  & motion detection & $1.00$ & $1.00$ & $1.00$ & $1.00$ & $1.00$ & $1.00$ & $1.00$ & $1.00$ & $1.00$\\
 \cline{1-12}
\multirow{8}{*}{lock}  &  \multirow{5}{*}{August Lock Pro} &  app opening & $0.78$ & $1.00$ & $0.78$ & $1.00$ & $1.00$ & $1.00$ & $0.77$ & $0.86$ & $0.88$\\
 &   & WiFi (un)locking & $0.80$ & $0.98$ & $0.82$ & $1.00$ & $1.00$ & $1.00$ & $0.75$ & $0.81$ & $0.91$\\
 &   & Bluetooth (un)locking & $0.90$ & $1.00$ & $0.90$ & $1.00$ & $1.00$ & $1.00$ & $0.99$ & $0.99$ & $1.00$\\
 &   & autolocking & $0.85$ & $1.00$ & $0.85$ & $1.00$ & $1.00$ & $1.00$ & $0.97$ & $1.00$ & $0.97$\\
  &   & manual (un)locking & $0.93$ & $1.00$ & $0.93$ & $1.00$ & $1.00$ & $1.00$ & $1.00$ & $1.00$ & $1.00$\\
  \cline{2-12}
  & \multirow{3}{*}{Schlage  WiFi  Deadbolt} &  WiFi (un)locking &  $0.89$ &  $1.00$ &  $0.89$ &  $1.00$ &  $1.00$ &  $1.00$ &  $1.00$ &  $1.00$ &  $1.00$\\
  &  & autolocking & $0.92$ & $1.00$ & $0.92$ & $1.00$ & $1.00$ & $1.00$ & $1.00$ & $1.00$ & $1.00$\\
  &  & manual (un)locking & $0.90$ & $1.00$ & $0.90$ & $1.00$ & $1.00$ & $1.00$ & $1.00$ & $1.00$ & $1.00$\\
\cline{1-12}
\multirow{6}{*}{plug}
  & \multirow{2}{*}{Amazon  Smart  Plug} &  on &  $1.00$ &  $1.00$ &  $1.00$ &  $1.00$ &  $1.00$ &  $1.00$ &  $1.00$ &  $1.00$ &  $1.00$\\
  & & off &  $0.96$ &  $1.00$ &  $0.96$ &  $1.00$ &  $1.00$ &  $1.00$ &  $1.00$ &  $1.00$ &  $1.00$\\
\cline{2-12}
  & Gosund WiFi Smart Socket & on or off  & $0.98$ & $1.00$ & $0.98$ & $1.00$ & $1.00$ & $1.00$ & $0.99$ & $0.99$ & $1.00$\\
\cline{2-12}
  & \multirow{2}{*}{TP-Link Plug}  & on  & $0.99$ & $1.00$ & $0.99$ & $1.00$ & $1.00$ & $1.00$ & $1.00$ & $1.00$ & $1.00$\\
  &  & off & $0.97$ & $1.00$ & $0.97$  & $1.00$ & $1.00$ & $1.00$  & $1.00$ & $1.00$ & $1.00$\\
 \cline{2-12}
 &  WeMo Plug & on or off & $0.98$ & $1.00$ & $0.98$ & $1.00$ & $1.00$ & $1.00$ & $0.99$ & $0.99$ & $1.00$\\

\bottomrule
\end{tabular}
\end{table*} 

From TABLE~\ref{tab:evaluate}, we observe that the performance of
IoTAthena depends on $r$.
For $r = 3$, IoTAthena achieves
a minimum accuracy of $0.78$,
a minimum precision of $0.98$,
and
a minimum recall of $0.78$.
When $r$ is increased to $11$, the performance of IoTAthena
improves,
with
precision of $1.00$,
accuracy of $0.99$ or better,
and recall of $0.99$ or better,
across all activities in our experiments.
When $r$ is further increased to $23$, the performance of IoTAthena
drops,
with a minimum accuracy of $0.75$,
a minimum precision of $0.81$,
and a minimum recall of $0.88$.
Based on this empirical evidence, we choose $r = 11$ as the
``optimal'' value for the tolerance parameter.

\begin{table}[htbp]
\scriptsize
\centering
\caption{Accuracy, precision, and recall of IoTAthena in the external MON(IOT)R dataset with $r$ set as $11$.}
\label{tab:IMC}
\begin{tabular}{l l l l l}
\toprule
 \textbf{Device Name} & \textbf{Activity} &  \textbf{$\mathcal{A}$} & \textbf{$\mathcal{P}$} & \textbf{$\mathcal{R}$}\\\hline
Amcrest Camera Wired & watch & $1.00$ &  $1.00$ &  $1.00$\\\hline
Blink Camera & watch & $1.00$ &  $1.00$ &  $1.00$\\\hline
Blink Security Hub & watch or photo & $1.00$ &  $1.00$ &  $1.00$\\\hline
Bulb1  & on or off & $1.00$ &  $1.00$ &  $1.00$\\\hline
Fire TV & menu & $0.95$ &  $1.00$ &  $0.95$\\\hline
Google Home Mini & volume or voice & $1.00$ &  $1.00$ &  $1.00$\\\hline
Insteon Hub & on or off & $1.00$ &  $1.00$ &  $1.00$\\\hline
Invoke & volume or voice & $1.00$ &  $1.00$ & $1.00$\\\hline
Lefun Camera Wired & watch or photo & $1.00$ &  $1.00$ &  $1.00$\\\hline
LG TV Wired & menu & $1.00$ &  $1.00$ &  $1.00$\\\hline
\multirow{2}{*}{Lightify Hub} & on or off & $1.00$ &  $1.00$ &  $1.00$\\
 & color & $1.00$ &  $1.00$ &  $1.00$\\\hline
Luohe Spycam & watch & $1.00$ &  $1.00$ &  $1.00$\\\hline
\multirow{2}{*}{Magichome Strip} & on & $0.98$ &  $0.98$ &  $1.00$\\
 & off & $1.00$ &  $1.00$ &  $1.00$\\\hline
Microseven Camera & watch & $1.00$ &  $1.00$ &  $1.00$\\\hline
Philips Bulb & on or off & $0.97$ &  $1.00$ &  $0.97$\\\hline
Samsungtv Wired & menu & $1.00$ &  $1.00$ &  $1.00$\\\hline
Sengled Hub & on or off & $0.98$ &  $1.00$ &  $0.98$\\\hline
Smartthings Hub & on or off & $1.00$ &  $1.00$ &  $1.00$\\\hline
T-philips Hub & on or off & $1.00$ &  $1.00$ &  $1.00$\\\hline
\multirow{4}{*}{TP Link Bulb} & on & $1.00$ &  $1.00$ &  $1.00$\\
 & off & $1.00$ &  $1.00$ &  $1.00$\\
 & color & $1.00$ &  $1.00$ &  $1.00$\\
 & dim & $1.00$ &  $1.00$ &  $1.00$\\\hline
 \multirow{2}{*}{TP Link Plug} & on & $1.00$ &  $1.00$ &  $1.00$\\
 & off & $1.00$ &  $1.00$ &  $1.00$\\\hline
  \multirow{2}{*}{Wink Hub2} & on & $1.00$ &  $1.00$ &  $1.00$\\
 & off & $1.00$ &  $1.00$ &  $1.00$\\\hline
 Xiaomi Hub & on or off & $0.98$ &  $0.98$ &  $1.00$\\\hline
 Xiaomi Strip & on or off & $1.00$ & $1.00$ &  $1.00$\\\hline
 Zmodo Doorbell & watch & $1.00$ & $1.00$ &  $1.00$\\
\bottomrule
\end{tabular}
\end{table}


We also evaluated the performance of IoTAthena using the MON(IOT)R dataset~\cite{Ren:IMC19}.
Due to the relatively small sample size
(between $30$ and $40$), we ran $4$-fold cross validation
instead of $6$-fold cross validation.
TABLE~\ref{tab:IMC} illustrates the accuracy, precision, and recall
measures of running IoTAthena against $25$ IoT devices in the MON(IOT)R dataset with $r$ set to $11$.
We observe that IoTAthena achieve
a minimum accuracy of $0.95$,
a minimum precision of $0.98$, and
a minimum recall of $0.95$.

The prior study~\cite{Trimananda:NDSS20} also evaluates the algorithm with 
the same the MON(IOT)R dataset~\cite{Ren:IMC19}. 
TABLE VI in~\cite{Trimananda:NDSS20} reports an average accuracy of 99.12\% on 3 IoT device activities 
for WAN Sniffer, and an average accuracy of 99.06\% on 
4 IoT device activities for WiFi sniffer. As shown in TABLE~\ref{tab:IMC}, 
our approach achieves an average accuracy of 99.57\% on 33 IoT device activities
on the same dataset. Therefore, our proposed IoTAthena system is able to generate 
signatures for more IoT device activities than~\cite{Trimananda:NDSS20} while achieving 
slightly better accuracy in identifying the signatures with the same public dataset.

In summary, experimental evaluations with our smart home testbed
and the MON(IOT)R dataset demonstrate that IoTAthena can
successfully unveil homogeneous IoT device activities from network
traffic logs.

\subsection{Performance of IoTAthena on Mixed Device Activities}
\label{sec:evalD}
\noindent
A significant benefit of our IoTAthena system lies in the realtime security
monitoring of IoT devices in smart homes, which has become an increasingly
important research topic.
Given the IoT network traffic logs, IoTAthena can accurately unveil the
sequence of IoT device activities over time and potentially detect
anomalous traffic patterns and behaviors towards IoT devices.

\begin{figure}[htbp]
\centering
\includegraphics[width=5.0in]{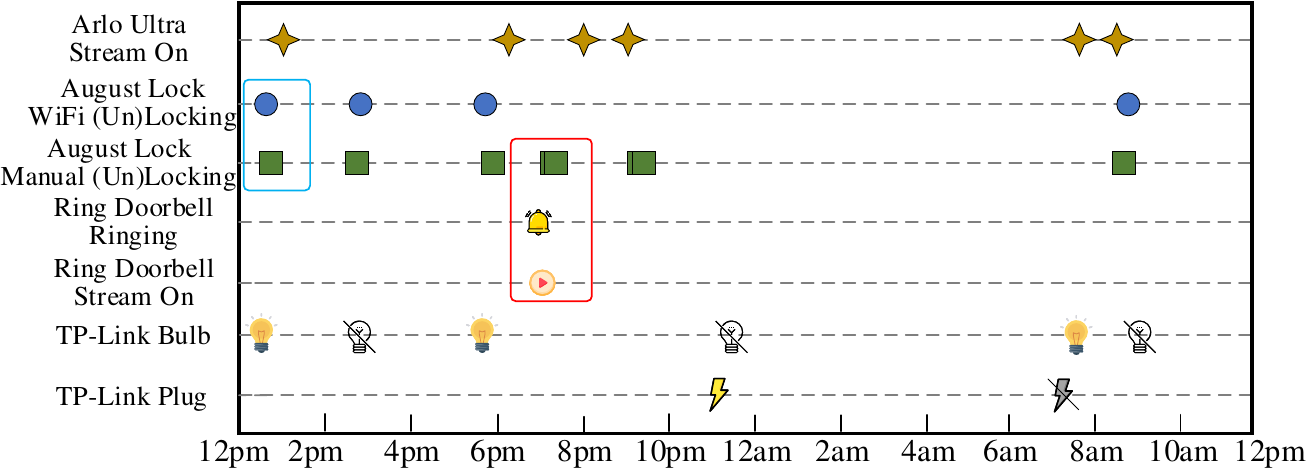}
\caption{IoT device activities discovered by IoTAthena in the small home environment during a 24-hour span.}
\label{fig:iotactivities}
\vspace{-0.15in}
\end{figure}

As a case study, we applied IoTAthena to unveil the activities of $5$ IoT devices in our smart home during a $24$-hour span.
The $5$ devices in this case study consist of Arlo Ultra Camera, August Lock, Ring Doorbell, TP-Link Bulb, and TP-Link Plug.
Fig.~\ref{fig:iotactivities} visualizes the time-series activities of these $5$ IoT devices discovered by IoTAthena during a $24$-hour time span in our smart home environment.

The two activities highlighted by the light blue box near the left end of Fig.~\ref{fig:iotactivities}
capture two consecutive user-triggered events at around 12:45pm:
\romannumeral1) (the homeowner) unlocked the August Lock with app
(from outside), indicated by the dark blue disk inside the light blue box,
and
\romannumeral2) manually locked the August Lock (after entering home),
indicated by the green square inside the light blue box.
Similarly, the four activities in highlighted by the red box at around 7:20pm in Fig.~\ref{fig:iotactivities}
reflect four consecutive events:
\romannumeral1) (a visitor) pressed the button on the ring doorbell,
which generated a push notification to the homeowner's smartphone;
\romannumeral2) (the homeowner) watched the video streaming feed on the ring
doorbell to check the visitor's identity;
\romannumeral3) (the homeowner) manually unlocked the August Lock to let the visitor in;
\romannumeral4) the August Lock was manually locked (from inside).

\begin{figure}[htbp]
\centering
\includegraphics[width=5.0in]{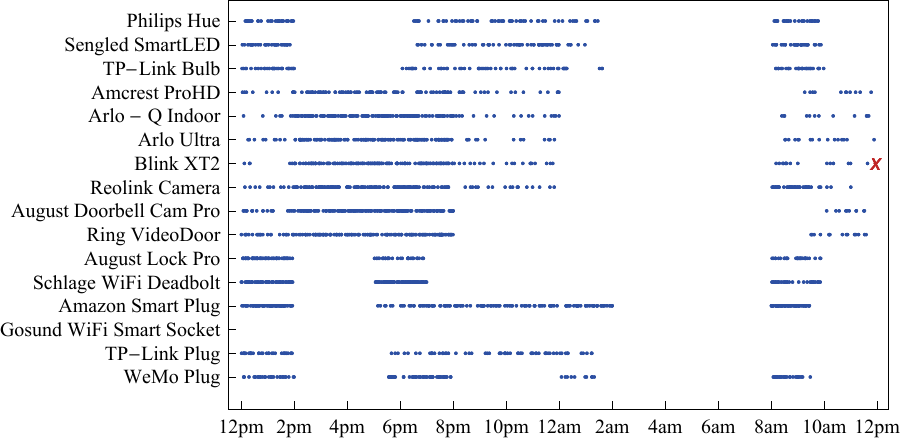}
\caption{Activity sequence extraction results of $16$ IoT devices in the smart home environment. The actual dates for running different IoT device activity experiments might vary, so the x-axis only denotes the time of the day from 12:00pm to 11:59am.}
\label{fig:sequence}
\vspace{-0.15in}
\end{figure}

To evaluate IoTAthena's ability in unveiling sequences of mixed IoT device
activities, we used IoTAthena to unveil the mixed IoT device activities of
all $16$ devices in our smart home from the network traffic,
in a $24$-hour span, from 12:00pm to 11:59am.
Fig.~\ref{fig:sequence} illustrates IoTAthena's performance
by comparing the ground truth activity sequences with the unveiled
activity sequences of $16$ IoT devices in the smart home testbed,
where a blue dot represents a successful match, while a red cross
represents a failed match.
The actual dates for running different IoT device activity experiments might
vary, so the x-axis only denotes the time of the day from 12:00pm to 11:59am.
As can be seen from the figure, IoTAthena correctly unveiled all but one
of the activities.
The only missed activity occurs with the Blink XT2 Camera.
Our root cause analysis revealed that IoTAthena missed one streaming activity
due to the unseen variation in packet length.

\vspace{5pt}
\noindent
In summary, our experimental evaluations based on a variety of heterogeneous IoT devices demonstrated that IoTAthena can effectively and accurately unveil individual IoT device activities as well as unveil IoT device activity sequences over time.
We note that our single smart home environment in the experiments has its own limitation in performing large scale experimental evaluations.
One of our future work is to deploy the IoTAthena system in a large number of smart homes to evaluate its performance and overhead.

\section{Conclusions and Future Work}
\label{sec:conc}
\noindent
This paper introduces {\em IoTAthena} to effectively and accurately unveil IoT
device activities from network traffic in smart homes.
We first recognize and  generate activity signatures of IoT device activities
consisting of ordered sequences of IP data packets by repeated and controlled
experiments.
\textcolor{black}{Subsequently, we design two polynomial time algorithms, {\tt sigMatch} and {\tt actExtract}.
The {\tt sigMatch} algorithm captures all matches of any given IoT device activity signature from real
network traffic logs. The {\tt actExtract} algorithm unveils the full activity sequences of all IoT
devices from the network traffic log.}
Through experimental evaluations based on a wide range of heterogeneous IoT
devices from a real smart home environment and a public IoT dataset, we
demonstrated that IoTAthena is able to accurately unveil IoT device activities
from raw network traffic logs.
We are in the process of designing and implementing a prototype system on commodity
home routers to evaluate the real-time feasibility of IoTAthena for unveiling IoT
device activities on the fly.
Another possible future work is to explore the benefits of IoTAthena in
detecting and mitigating security threats towards vulnerable IoT devices.


\end{document}